\documentclass[twocolumn,american,nofootinbib]{revtex4-2}
\usepackage{amsmath}
\usepackage[T1]{fontenc}
\usepackage[utf8]{inputenc}
\usepackage{colortbl}
\usepackage{babel}
\usepackage{array}
\usepackage{booktabs}
\usepackage{varwidth}
\usepackage{graphicx}
\usepackage{subfig}
\usepackage{setspace}
\usepackage[pdfusetitle,
 bookmarks=true,bookmarksnumbered=false,bookmarksopen=false,
 breaklinks=false,pdfborder={0 0 1},backref=false,colorlinks=true]
 {hyperref}
\hypersetup{
 urlcolor=blue, citecolor=blue, linkcolor=blue}

\makeatletter

\providecommand{\tabularnewline}{\\}
\newenvironment{cellvarwidth}[1][t]
    {\begin{varwidth}[#1]{\linewidth}}
    {\@finalstrut\@arstrutbox\end{varwidth}}

\usepackage{soul}
\usepackage{xcolor}

\makeatother

\begin{document}
\title{Anisotropic truncation for turbulent transport and zonal flows in the Hasegawa-Wakatani
system}
\author{P. L. Guillon$^{1,2}$, R. Angles$^{1}$, Y. Sarazin$^{3}$ and Ö.
D. Gürcan$^{1}$}
\affiliation{$^{1}$Laboratoire de Physique des Plasmas, CNRS, Ecole Polytechnique,
F-91120 Palaiseau, France}
\affiliation{$^{2}$ENPC, Institut Polytechnique de Paris, 77455 Champs-sur-Marne,
Marne-la-Vallée cedex 2, France}
\affiliation{$^{3}$CEA, IRFM, F-13108 Saint-Paul-lez-Durance, France}
\begin{abstract}
Reduced models based on an anisotropic truncation of the Fourier space,
retaining only a few poloidal wave-numbers while keeping the full
radial resolution, are developed and first applied to the Hasegawa-Wakatani
system. The impact of the truncation is studied first by considering
the fixed-gradient formulation, and by comparing to direct numerical
simulations (DNS). The turbulent particle flux, and the transition
from quasi-two dimensional turbulence to the zonal flow (ZF) dominated
state, are used as the main criteria for validation. It is found that
at least 4 poloidal modes, distributed around the most unstable mode,
are needed to observe a sharp transition, and that about 10 modes
are needed to reproduce each transport regime. Then, similar reduced
models are developed in a flux-driven formulation and compared to
the DNS, focusing on two cases far from the nonlinear threshold of
the transition from turbulence to zonal dominated states of the fixed
gradient formulation. In that case, using 10 modes allows to match
the probability distribution function of the particle flux of the
DNS approximately. Considering the role played by different poloidal
scales in the turbulent cascade, it is observed that in the turbulent
state, an inverse energy cascade in radial wave-numbers takes place
at large poloidal scales, while a forward enstrophy cascade in radial
wave-numbers is observed to occur at smaller poloidal scales. Moreover,
when they form, ZFs feed on poloidal scales that are around and slightly
smaller than the injection scale, while giving their energy to the
larger poloidal scales. In that case, there is an anisotropic inverse
energy transfer, akin to inverse cascade, from the energy injection
to the large poloidal scales \emph{through} ZFs, while the forward
enstrophy cascade seems to stay isotropic.
\end{abstract}
\maketitle

\section{Introduction}

\label{sec:Introduction}

In this work, we develop a certain class of reduced models, that we
call Poloidally Truncated Models (PTMs), to investigate transport
in instability-driven turbulent models for tokamak plasmas. These
models are constructed by keeping the complete resolution in the radial
direction, while keeping only a few poloidal modes in Fourier space,
and thus run much faster (nearly 20 times) than a fully resolved direct
numerical simulation (DNS). This strategy can be applied to both fixed-gradient
and flux-driven formulations of a system describing plasma turbulence.
In the latter case, it provides a ``1.5D'' transport model where the
turbulent fluxes are self-consistently computed from the few poloidal
modes that are retained.

As a proof of concept, in this article, we apply these models to the
Hasegawa-Wakatani (HW) system \citep{hasegawa:1983}, which we consider
as the absolute minimum, non-trivial, model of plasma turbulence,
with drift-waves, driven by a linear instability, and that can form
zonal flows (ZFs) \citep{numata:2007,guillon:2025,guillon:2025d}.
However, the eventual goal is to generalise this strategy to various
more realistic systems, including interchange, fluid ion temperature
gradient (ITG) turbulence, gyrofluid models, and potentially gyrokinetic
simulations.

The motivation for performing such an anisotropic reduction is threefold.
First, the \emph{raison d'être} of turbulence simulations, be it fluid
or gyrokinetics, is to compute turbulent transport in the radial direction
either in the form of anomalous diffusion coefficients, or equivalently
the turbulent fluxes in fixed gradient formulations, given the gradients.
On the other hand, in flux-driven formulations \citep{garbet:1998,gillot:2023},
turbulent transport is studied in the form of self-consistently evolving
radial profiles, given the sources and sinks, driving an average flux
through the system. Secondly, many of the non-trivial turbulent phenomena,
such as ZFs, transport barriers or avalanches \citep{biglari:1990,hinton:1991,hinton:1993,diamond:1995,carreras:1996,diamond:2001,diamond:2005,dif-pradalier:2015,gurcan:2015,cao:2025,diamond:2025},
which are of great interest to profile self-organisation and hence
the confinement, are mostly meaningful from a flux surface averaged
perspective. Reduced model to a 1D system \citep{panico:2025,panico:2025a},
to a few Fourier modes \citep{terry:1983,gurcan:2022,guillon:2025},
or even to 0D dynamical systems \citep{diamond:1994,kim:2003,araki:2023,kosuga:2025},
also provide insights to understand the complex interplay between
turbulence and its large-scale self-organisation. Finally, the energy
injection through linear instability, and the initial non-linear couplings,
take place mainly at large scales, close to the most unstable wave-number
\citep{horton:1999}, and it is mostly those large scales that contribute
to turbulent fluxes.

Hence, while it is important to keep a decent resolution in the radial
direction, the number of poloidal modes can, in principle, be reduced
substantially, especially if one is interested in representing the
mechanisms of energy injection, some non-linear couplings, and the
possibility of self-organisation, resulting in transport barriers,
staircases and avalanches. Finding the appropriate minimal set of
poloidal modes that achieves this, is the goal of this article.

PTMs can be seen as generalisations of the Tokam1D model \citep{panico:2025,panico:2025a}
for flux-driven plasma turbulence, that keeps only one poloidal mode
and the full radial resolution, in a physical system similar to the
Hasegawa-Wakatani model (which also includes interchange drive and
the diamagnetic non-linearity \citep{sarazin:2021}). Tokam1D itself,
can be considered to be related to the quasilinear models \citep{marston:2016},
commonly used in the geophysical fluid dynamics (GFD) context, but
here, they are obtained by decimating the Fourier space instead of
the interaction kernel. Such models ignore turbulence-turbulence interactions
altogether, which may be reasonable in the ZF dominated regime, but,
these interactions may be needed to correctly describe the self-organisation
of the system, as well as the transition from a fully turbulent state
to a ZF dominated regime \citep{numata:2007,guillon:2025}.

In this spirit, the PTMs are similarly related to what are called
``generalised quasi-linear models'' in the GFD community \citep{marston:2016,nivarti:2024a},
again, obtained by a truncation of the Fourier space \citep{gurcan:2023}
to a small number of large scale modes, rather than a truncation of
the convolution operator to include only the interactions among those
modes. This is justified for the plasma problem, since the main interest
in turbulence in magnetised plasmas is the transport that it generates.
Hence, if keeping a few large scale modes is sufficient to reproduce
the mean transport dynamics and some of its statistical events, then
the small scales can be dropped for the sake of computing self-consistent
fast estimates of the turbulent transport.

For the truncations where the large scales are not sparsified, this
corresponds to a Large Eddy Simulation (LES) \citep{sagaut:2006,lesieur:2008}
of the system, with an anisotropic filter of cutoff wave-number ${\bf k}^{c}=(k^{c}_{x},k^{c}_{y})$
such that $k^{c}_{x}=k^{DNS}_{max},k^{c}_{y}\ll k^{DNS}_{max}$, with
$k^{DNS}_{max}$ the largest wave-number available in the DNS, that
we call ``poloidal'' LES (pLES). The main difference is that, in
PTMs, in order to lower the number of modes retained, we usually also
remove some large scale poloidal modes from the LES grid, and centre
the poloidal modes around the most unstable mode.

In short, PTMs generalise models such as Tokam1D to include multiple
poloidal modes, and thus also interactions among those, but they may
still drop both large and small poloidal scales. In contrast pLES
(which is a particular case of the PTM) drops only small poloidal
scales, keeping all the large scales of the DNS.

Such a truncation should eventually be compensated by a closure term,
but due to the anisotropic nature of the problem, and of the non-local
interactions between turbulence and ZFs \citep{krommes:2000}, it
is not obvious what precise closure to implement. In this paper, which
represents a first attempt, we choose not to apply any closure terms,
partly to see the limitations of such an approach, and we use instead
the same diffusion coefficients as in the DNS. This yields already
promising results. Our explanation is that, since the full resolution
in the radial direction is retained, and the non-linear term eventually
couples $k_{x}=0$ modes to high $k_{x}$ modes, where the regular
dissipation acts, this is possibly enough to provide sufficient dissipation
to large poloidal scales, where the dynamics is relevant for transport
and zonal flow formation. However, we plan in the future to explore
what kind of closure terms can be added, similar to what is used in
regular LES, e.g. the dynamical procedure developed by Germano et
al. \citep{germano:1991}, which has been applied to gyrokinetic simulations
by Morel et al. \citep{morel:2012}. Note that, since the resulting
turbulence is anisotropic, and modulated by ZFs (not really homogeneous),
the main challenge in such a closure is the proper treatment of subgrid
contribution to ZF damping, as well as the effect of ZFs on turbulent
eddy damping. Both of these processes are rather non-trivial to model.

In terms of implementation, numerical simulations of PTMs can be performed
using \emph{the same} pseudo-spectral solver as the DNS, but simply
reducing the resolution along the poloidal direction, and adjusting
the box size to select the desired modes. Due to this construction,
the poloidal modes are always spaced regularly, which results in a
configuration where triadic interactions are naturally permitted among
all the modes, as compared to an arbitrary selection of modes. Then,
since we always include the poloidal wave-number corresponding to
the most unstable mode, in order to retain the correct mechanism for
the linear instability, all PTMs that we use have a small prescribed
number of poloidal modes distributed around the most unstable mode.
What we note is that models that display the best performance in
reproducing the DNS are those for which modes are distributed at scales
both larger and smaller than the most unstable mode $k_{y0}$, compared
to including only larger or smaller poloidal scales. This may be due
to the nature of the non-linear transfers and interactions with ZFs
induced by these different regions, as discussed in Section \ref{sec:pollines}.
A similar construction leads to flux-driven PTMs, using the newly-developed
P-FLARE \citep{guillon:2025a,guillon:2025c} code, which performs
pseudo-spectral simulations of a flux-driven fluid system using the
penalisation method \citep{angot:1999}.

The Hasegawa-Wakatani (HW) system constitute an interesting test-bed
for a proof of concept, because it exhibits a sharp transition between
a turbulence state, and a state dominated by ZFs \citep{numata:2007}.
Such a transition between two different transport regimes is reminiscent
of the Dimits shift observed in earlier gyrokinetic and fluid simulations
\cite{dimits:2000} and thus provide an important challenge for a
reduced model to be able to reproduce. The interplay between turbulence
and ZFs, the level of which is given in the HW system by the fraction
of energy in ZFs $\Xi_{\mathcal{K}}$, is relevant for L-mode plasmas,
and may have a key role in the L-H transition, for example in the
limit cycle oscillations \cite{kim:2003}, which have been observed
experimentally in several machines \cite{schmitz:2012,hillesheim:2016,cavedon:2016}.
Having a reduced transport model such as PTMs, which describes self-consistently
the interaction between ZFs and turbulence, can potentially be key
for transport modelling and profile prediction, which can then be
compared to experiments. Note that the HW model also allows to address
the role of cross-phase between electrostatic potential and density
fluctuations on the transport \cite{an:2017,leconte:2021}.

Moreover, the HW system has a subcritical zonal bifurcation around
the transition point, with a hysteresis loop \cite{grander:2024,guillon:2025,guillon:2025d},
following the behaviour that has also been observed in trapped particle
simulations \cite{gravier:2017} and more recently around the Dimits
threshold in ITG \cite{marquant:2026}. In fixed-gradient simulations,
this may manifest as a system that can jump between two transport
regimes, as observed in gyrokinetic simulations \cite{peeters:2016}.
In flux-driven system, the hysteresis may be a source of profile stiffness,
where different level of fluxes (corresponding to different levels
of zonal flows) could be obtained with the same mean profiles \cite{guillon:2025c}.
Therefore, a simple model, implemented in such a way that one can
easily switch between flux-driven and fixed-gradient formulations,
may provide an explanation, and/or a simple framework to study the
profile stiffness observed in experiments \cite{stober:2000,wolf:2003,garbet:2004,mantica:2009,ryter:2011}.

Applying PTMs to the fixed-gradient Hasegawa-Wakatani (HW) system,
and using the transition between 2D turbulence and ZFs as the main
validation criterion, we find that at least 4 poloidal modes, equally
distributed around the most unstable mode $k_{y0}$, are needed in
order to observe a sharp transition, although the ZF fraction is overestimated
in the turbulent regime. With more than 10 modes, both turbulent and
ZF dominated states are well described, and the particle flux scaling
with the linear parameters is reproduced, but we find that the transition
point is systematically shifted. The hysteresis is also not captured
by these models in their current form. One may speculate that they
can be recovered by a ZF level dependent dissipation, since the actual
transition point seems to be very sensitive to the diffusion coefficients.
In fact, matching the hysteresis curve could probably be used to design
such a closure. However, this requires a separate detailed study,
which we leave to future works.

Then, turning to the particle flux-driven HW system \citep{guillon:2025c},
we obtain similar ZF levels and track the time evolution of the mean
density gradient observed in the DNS, if we use at least the 4 poloidal
modes that are equally distributed around the most unstable mode.
Furthermore, using 10 modes, we can reproduce the probability distribution
function of the particle flux, rather closely, which means that the
PTMs can even describe part of the statistical properties of the flux-driven
system. However, the profile stiffness behaviour is not yet captured,
which we expect since the fixed-gradient system does not have the
hysteresis.

The models can describe avalanches, if defined as profile relaxation
events, and more strongly reduced models may even favour them a bit
too much, as seen in the study of their statistics with different
levels of reductions, in comparison to the DNS. However, note that
the HW system do not drive a lot of streamers, except maybe at the
linear phase, so if one is interested in studying the effect of streamers
as a separate class of modes, one should either use interchange or
consider for example fluid ITG or electron temperature gradient driven
(ETG) mode, where the streamers may appear as nonlinear structures
\citep{gurcan:2004,gurcan:2004a}.

Finally, to justify the selection of the poloidal wave-numbers, we
investigate, the role played by each poloidal mode $k_{y}$ and its
associated side-bands $\left(k_{x},k_{y}\right)$, where $k_{x}$
is the radial wave-number, in the turbulent energy and enstrophy transfers
\citep{camargo:1995}. Particularly, we find that, in the turbulent
state, poloidal scales larger than the most unstable mode, i.e. $k_{y}<k_{y0}$,
mainly cause inverse energy transfer along $k_{x}$, in line with
the tendency of inverse energy cascade in 2D turbulence, while the
modes with $k_{y}>k_{y0}$ mainly generate direct energy transfer
along $k_{x}$, through the enstrophy cascade. In contrast, in the
ZF dominated state, we observe that the stationary state is achieved
as a balance between the largest poloidal scales (i.e. small $k_{y}<k_{y0}/2$)
taking energy from ZFs, and poloidal scales that are slightly smaller
scale than the most unstable mode (i.e. $k_{y}\gtrsim k_{y0}$) feeding
ZFs. That case is also consistent with a dual cascade, where ZFs induce
an anisotropic energy transfer from the injection scale to larger
poloidal scales, akin to inverse cascade, while the direct enstrophy
cascade seems to remain isotropic. It is thus important to include
both regions $k_{y}<k_{y0}$ and $k_{y}>k_{y0}$ in the truncated
models in order to reproduce the non-linear transfers observed in
the DNS.

The rest of the article is organised as follows. In Section \ref{sec:pres},
the construction of the PTMs studied in the article is presented,
after introducing the Hasegawa-Wakatani model. The reduced models
are compared to fixed gradient DNS by performing scans of the adiabaticity
parameter in Section \ref{sec:fixedgrad}. Then in Section \ref{sec:fluxdriven},
the models are applied to the flux-driven system, with a particle
source, in both turbulence and zonal flow dominated regimes. Finally,
the role of different poloidal modes in the turbulent energy and enstrophy
transfers are investigated in Section \ref{sec:pollines}.

\section{Hasegawa-Wakatani system and poloidally truncated models set-up}

\label{sec:pres}

First, we present both fixed gradient and flux-driven Hasegawa-Wakatani
systems, and define some validation criteria for the reduced models,
based on known properties of the full system, which are, to some extent,
common to most instability-driven turbulent systems in tokamaks plasmas.
Then, we construct PTMs, which in this case are based on the linear
properties of the Hasegawa-Wakatani system.

\subsection{The Hasegawa-Wakatani system}

\subsubsection{The classic, fixed gradient system}

The Hasegawa-Wakatani (HW) is a minimal model of plasma turbulence,
which describes instability-driven turbulence and its self-organisation
into zonal flows (ZF) in a 2D plane $(\mathbf{e}_{x},\mathbf{e}_{y})$
of size $(L_{x},L_{y})$ orthogonal to a constant and uniform magnetic
field $\mathbf{B}=B\mathbf{e}_{z}$. Here, the spatial variables $x$
and $y$ are normalised to $\rho_{s}=c_{s}/\Omega_{i}$ the sound
Larmor radius and time $t$ is normalised to $\Omega_{i}=eB/m_{i}$
the cyclotron frequency, where $c_{s}=\sqrt{T_{e}/m_{i}}$ is the
plasma sound speed, with $T_{e}$ the electron temperature (assumed
constant) and $m_{i}$ the ion mass. The $x$ direction corresponds
to the radial direction in a tokamak, while $y$ corresponds to the
poloidal direction. A linear background density profile $n_{r0}(x)$,
with a constant density gradient $\kappa\equiv-(1/n_{r0})(dn_{r0}/dx)$,
is imposed, which acts as the free energy source for the dissipative
drift-wave instability. The modified HW equations are \citep{hasegawa:1983,numata:2007,guillon:2025}
\begin{subequations}
\label{eq:hw} 
\begin{align}
\partial_{t}\nabla{^{2}}\phi+[\phi,\nabla{^{2}}\phi] & =C(\widetilde{\phi}-\widetilde{n})+\nu\nabla^{4}\widetilde{\phi},\label{eq:hw1}\\
\partial_{t}n+[\phi,n]+\kappa\partial_{y}\phi & =C(\widetilde{\phi}-\widetilde{n})+D\nabla^{2}\widetilde{n},\label{eq:hw2}
\end{align}
\end{subequations}
 where $\phi$ is the electric potential normalised to $T/e$, with
$T$ the electron temperature and $e$ the unit charge, $\nabla^{2}\phi$
is the vorticity and $n$ is the density perturbation, normalised
to a background reference density $n_{0}$. The Poisson bracket, defined
as $[\phi,A]\equiv\partial_{x}\phi\partial_{y}A-\partial_{y}\phi\partial_{x}A$,
corresponds to the advection of the field $A$ by the $E\times B$
drift velocity $\mathbf{v}_{E\times B}\equiv-\partial_{y}\phi\mathbf{e}_{x}+\partial_{x}\phi\mathbf{e}_{y}$.
The two equations are coupled by the adiabaticity parameter $C=v^{2}_{te}k^{2}_{\parallel}/\left(\nu_{c}\Omega_{i}\right)$,
with $k_{||}$ the wave-number of fluctuations parallel to the magnetic
field lines, $\nu_{c}$ the collision frequency and $v_{te}=\sqrt{T_{e}/m_{e}}$
the electron thermal velocity, where $m_{e}$ is the electron mass.
Note that we decomposed perturbations $A$ into their zonal and non-zonal
parts: $A=\overline{A}+\widetilde{A}$, where $\overline{A}\equiv\langle A\rangle_{y}=\frac{1}{L_{y}}\int^{L_{y}}_{0}Ady$\emph{
}is averaged along the poloidal direction, so that $\langle\widetilde{A}\rangle_{y}=0$.
Small scale dissipation is introduced on non-zonal fluctuations by
the viscosity and diffusion terms, with the coefficients $\nu$ and
$D$, respectively. We define the total mean kinetic energy as $\mathcal{K}\equiv\frac{1}{2L_{x}L_{y}}\int\int|\mathbf{v}_{E\times B}|^{2}\,dxdy$,
and its zonal part as $\mathcal{K}_{ZF}\equiv\frac{1}{2L_{x}}\int(\partial_{x}\overline{\phi})^{2}\,dx$.

The value of the ratio $C/\kappa$ determines the non-linear regime
of the system \citep{numata:2007}. In well resolved DNS studies,
it was observed that for $C/\kappa<0.1$, the system behaves like
2D hydrodynamic turbulence, while for $C/\kappa>0.1$, it is dominated
by ZFs and becomes quasi-1D. Using $C/\kappa$ as the control parameter,
and the ratio of zonal to total kinetic energy $\Xi_{\mathcal{K}}=\mathcal{K}_{ZF}/\mathcal{K}$
as the order parameter, there is an abrupt transition around the transition
point $C/\kappa\approx0.1$, suggesting a 1\textsuperscript{st} order
phase transition. One can also observe a hysteresis loop around the
critical point, such that once they have formed, ZFs can exist even
below the critical point, and that something analogue to latent heat
is required to collapse them \citep{guillon:2025}.

The radial particle flux $\Gamma\equiv\langle\widetilde{n}\widetilde{v_{x}}\rangle_{xy}=-\langle\widetilde{n}\partial_{x}\widetilde{\phi}\rangle_{xy}$,
averaged over the 2D domain, is the quantity of key interest for turbulent
transport in this system. It has been observed \citep{hu:1997,guillon:2025}
that the particle flux decreases as a function of the control parameter
as $\left(C/\kappa\right)^{\alpha}$ where $\alpha=-1/3$ in the turbulent
regime $\left(C/\kappa<0.1\right)$, and much steeper, with $\alpha=-2$,
when ZFs dominate $\left(C/\kappa>0.1\right)$, possibly as a consequence
of shear suppression \citep{biglari:1990}. This scaling can be captured
using a simple quasi-linear estimate normalised by the factor $\left(1-\Xi_{\mathcal{K}}\right)$
\citep{guillon:2025}, since only non-zonal fluctuations contribute
to $\Gamma$.

Both features of the transition, i.e. i) the sharp transition of the
ZF level $\Xi_{\mathcal{K}}=\mathcal{K}_{ZF}/\mathcal{K}$, and ii)
the double power law scaling of the particle flux with $C/\kappa$,
can serve as the validation criteria for any reduced model of the
HW turbulence, since they are both related to the level of turbulent
transport and the existence of self-generated ZFs acting as local
transport barriers for this system. A third criterion may be to obtain
the same transition point as the DNS, however we observe systematic
shifts of the threshold, depending on the mode selection, which calls
for further investigations on the underlining mechanisms of the transition.
The last criterion could be the ability of the model to reproduce
the hysteresis, but this requires better statistics and is significantly
more complicated to reproduce with reduced models, which we think
may be achieved with a ZF level dependent subgrid viscosity that we
leave for future work.

\subsubsection{The flux-driven system}

The system represented by Eq. \ref{eq:hw} corresponds to a local,
fixed gradient formulation of the Hasegawa-Wakatani model, in that
the mean density gradient $\kappa$ is imposed, and is not evolved
in response to the turbulent particle flux that it causes, even though
local corrugations of density may develop. In order to represent the
interplay between turbulence and transport correctly, one should instead
consider the particle flux-driven formulation \citep{guillon:2025c},
which can be achieved by replacing the zonal part of the density equation
of Eq. \ref{eq:hw2} with a transport equation on the full radial
density profile $n_{r}\equiv-\kappa(t)(x-L_{x})+\overline{n}(x,t)$:
\begin{equation}
\partial_{t}n_{r}+\partial_{x}\Gamma_{n}=S_{n}(x,t),\label{eq:nrfluxdriven}
\end{equation}
where $\Gamma_{n}(x)\equiv\langle\widetilde{n}\widetilde{v_{x}}\rangle_{y}$
and $S_{n}$ is a particle source term. Note that we could also introduce
a ``neoclassical'' diffusion term, say of the form $D_{0}\partial^{2}_{x}n_{r}$.
Furthermore, one may also use a similar equation for the radial profile
of the poloidal velocity $u_{E\times B}\equiv\langle\partial_{x}\phi\rangle_{y}$,
in order to include a mean $E\times B$ flow and consider its relaxation
towards a neoclassical profile for a more complete turbulent transport
model \citep{hinton:1991,chone:2015}. Here, since our goal is to
study if the reduction is faithful to the original model, we use a
model with only zonal flows, and no $E\times B$ velocity profile.

We require that the reduced models correctly evolve the ZF level and
the density profile, so that the mean gradient at steady-state is
such that the mean flux balances the source. Additionally, we require
that PTMs reproduce the particle flux $\Gamma_{n}$ probability distribution
function (PDF) correctly, in order to accurately describe the statistics
of the stationary state.

\subsection{Set-up of the poloidally truncated models}

Here, we consider the construction of the PTMs, which are based on
the linear properties of the HW system.

\subsubsection{Linear properties of the Hasegawa-Wakatani system}

Linearising Eqs. \ref{eq:hw}, and taking the spatial Fourier transform,
we obtain the dispersion relation of the linear system, which gives
two eigenvalues 
\begin{equation}
\omega^{\pm}_{k}=\omega^{\pm}_{k,r}+i\gamma^{\pm}_{k},\label{eq:HWeigen}
\end{equation}
where $\omega^{\pm}_{k,r}$ are the real frequencies, and $\gamma^{\pm}_{k}$
are the growth ($+$ modes) or damping ($-$ modes) rates \citep{gurcan:2022}.
The analytical expression for $\omega^{\pm}_{k}$ is given in Appendix
\ref{sec:app-eigenvalues}. Note that in the hydrodynamic limit $C\ll1$,
the growth rate reduces to $\gamma^{+}_{k}\approx\sqrt{C\omega^{HM}_{k}(1+k^{2})/(2k^{2})}$,
while in the adiabatic limit $C\gg1$, it becomes $\gamma^{+}_{k}\approx(1/C)(\omega^{HM}_{k})^{2}k^{2}/(1+k^{2})$,
where $\omega^{HM}_{k}=\kappa k_{y}/(1+k^{2})$ is the Hasegawa-Mima
frequency.

\begin{figure}[h!]
\centering{}\includegraphics[width=.8\columnwidth]{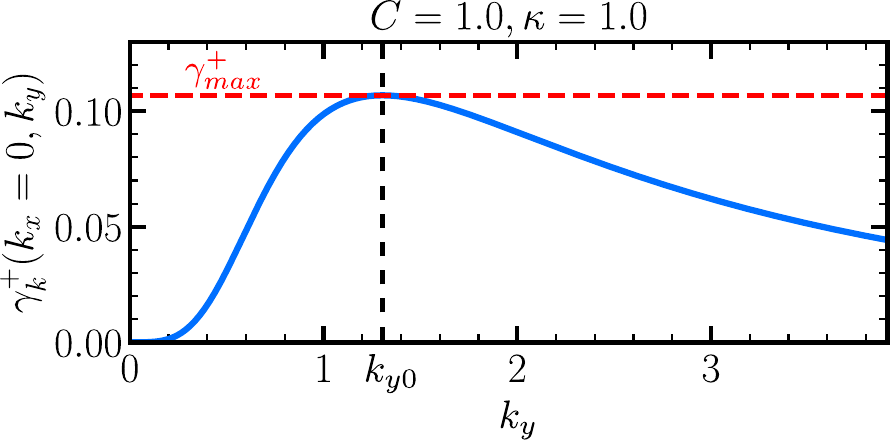}\caption{Linear growth rate $\gamma^{+}_{k}$ of the HW system as a function
of the poloidal wave-number $k_{y}$ with $k_{x}=0$, $C=1$ and $\kappa=1$
(in the inviscid case $\nu=D=0$). Dashed lines show the maximum growth
rate $\gamma^{+}_{max}$ and the associated most unstable wave-number
$k_{y0}$.}
\label{fig:gamky}
\end{figure}

For instability-driven turbulence, it is the growth rate $\gamma^{+}_{k}$
of the unstable modes that corresponds to the (linear) energy injection
mechanism in the system. For the HW system, this growth rate is symmetric
in $k_{x}$ and maximum for a purely poloidal wave-number $k_{y0}$,
because the background density gradient $\kappa$ is in the radial,
i.e. $x$ direction. More generally, for drift instabilities, it is
the background gradient of potential vorticity $dq_{0}/dx$, that
provides the free energy source, while the non-adiabatic electron
response allows the system to tap it \citep{horton:1999,gurcan:2015,gurcan:2024,guillon:2025d}.

The growth rate $\gamma^{+}_{k}$ as a function of $k_{y}$, with
$k_{x}=0$, $C=1$ and $\kappa=1$ (in the inviscid case $\nu=D=0$)
is shown in Figure \ref{fig:gamky}, where we denote the most unstable
wave-number $k_{y0}$ and the corresponding maximum growth rate $\gamma^{+}_{max}$
by dashed lines. These two quantities roughly correspond to energy
injection scale and the energy injection rate respectively. Both vary
with $C$ as can be seen in Figure \ref{fig:gamky0C} of the Appendix
\ref{sec:app-eigenvalues}.

\subsubsection{Construction of the PTMs}

Poloidally truncated models are obtained by projecting the Fourier
space grid of a DNS on a grid that retains only a few modes in the
poloidal direction, centred around the most unstable mode, along with
the full radial resolution. For a DNS with a typical padded resolution
of $N_{px}\times N_{py}=1024^{2}$, a PTM with $n_{y}=10$ poloidal
modes would correspond to a resolution of $N_{px}\times N'_{py}=1024\times34$
(i.e. $\lfloor34/3\rfloor-1=10$ with $2/3$ padding and Hermitian
symmetry), with a box size that is adjusted so that the most unstable
mode matches the central poloidal wave-number of the unpadded part.
The main idea is to coarsen the poloidal resolution in order to retain
the energy injection by the linear instability, and the self-organisation
mechanisms of turbulence in coherent structures such as ZFs and streamers.
Therefore, we use the most unstable wave-number $k_{y0}$ as a reference
to select which poloidal modes we keep.

\begin{figure}[htbp]
\centering{}\includegraphics[width=1\columnwidth]{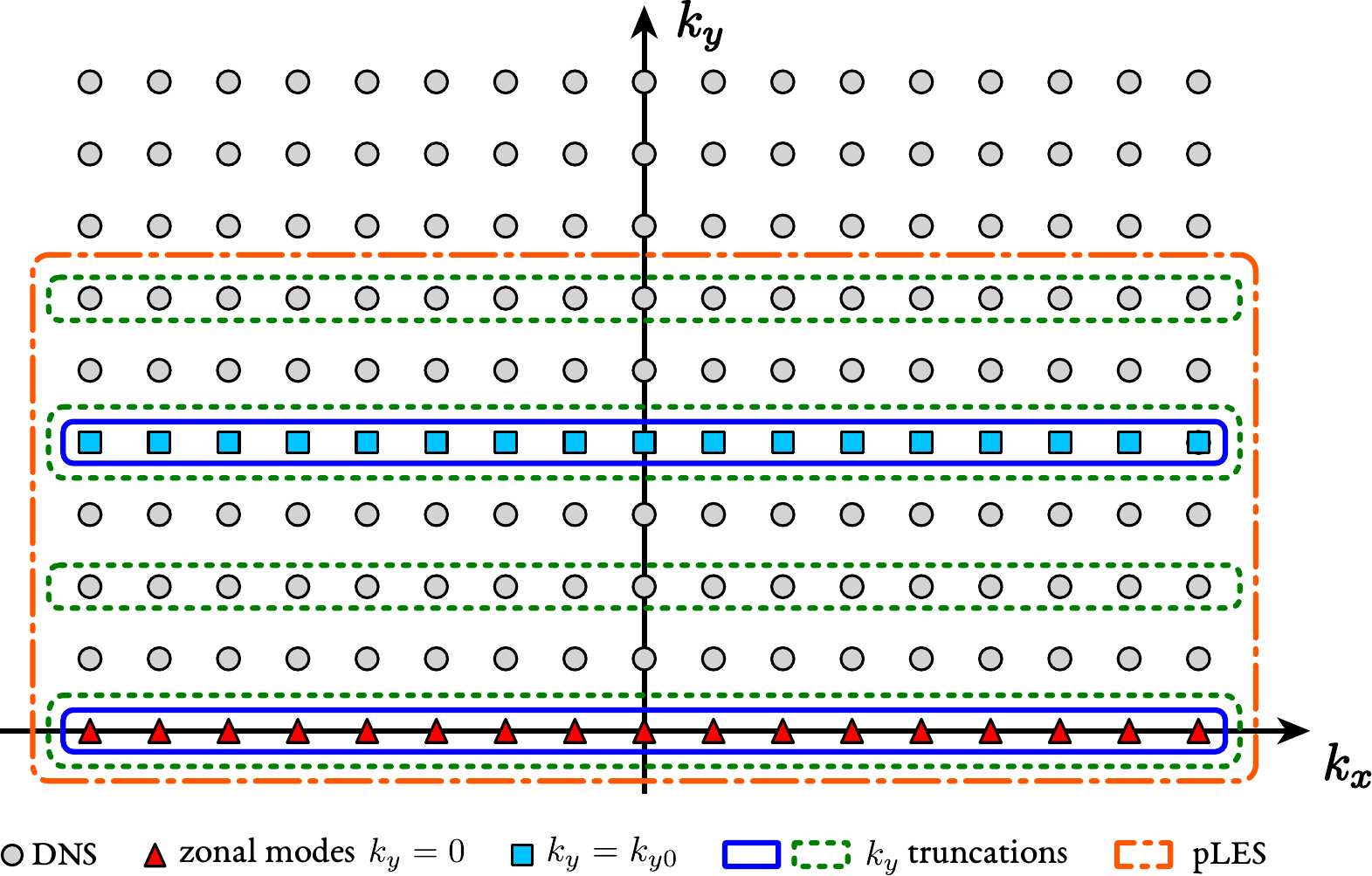}\caption{Schematic of the modes kept in PTMs, compared to a matching box DNS
Fourier grid (grey dots). The modes with poloidal wave-number equal
to the most unstable mode $k_{y0}$ are shown in blue squares. Zonal
modes (red triangles, $k_{y}=0$) are always kept. Blue, green and
orange: PTMs with 1, 3 and 6 poloidal modes kept. The latter is a
LES along the poloidal direction (pLES).}
\label{fig:PTRM_scheme}
\end{figure}

An illustration of PTMs is given in Figure \ref{fig:PTRM_scheme},
where several reductions are compared to a DNS grid (grey dots). The
zonal modes (red triangles, with $k_{y}=0$) are always kept. Keeping
only one poloidal mode (blue frame) corresponds to the Tokam1D model
\citep{panico:2025,panico:2025a}, and here we take the most unstable
mode $k_{y0}$ (blue squares). We can then add a few large poloidal
scales close to $k_{y0}$ (green and orange frames). Keeping all the
large scale poloidal modes from DNS (orange) corresponds to performing
an anisotropic LES in the poloidal direction (pLES). Note that keeping
the large $k_{x}$ modes allows to dissipate the energy at high wave-numbers.
Therefore, it seems not necessary to introduce complicated closure
schemes to compensate the truncation in the poloidal direction.

In the following, we construct PTMs by specifying their poloidal resolution
$n_{y}$ and the selected wave-numbers. Since $\phi$ and $n$ are
real, their Fourier transform are Hermitian symmetric, i.e. $\phi_{-k},n_{-k}=\phi^{*}_{k},n^{*}_{k}$,
where $*$ denotes complex conjugation. Therefore, in Fourier space,
we only need to specify which $k_{y}>0$ modes are kept. Hence, $n_{y}$
corresponds in fact to the number of modes --- or resolution ---
of the half poloidal axis $k_{y}>0$.

Note that in PTMs, we drop an important fraction of the poloidal modes,
but keep \emph{all} the non-linear interactions implied by the Fourier
modes that are kept, in contrast to the generalised quasi-linear models
\citep{marston:2016,nivarti:2024a}, where all modes are kept but
a large fraction of the interactions are dropped. Note also that using
a regular grid of a pseudo-spectral code, we naturally ensure the
existence of many triadic interactions, compared to choosing arbitrary
$k_{y}$ modes.

\subsubsection{The parameters of the Reduced models}

\label{subsec:setup}

To investigate the ability of PTMs to reproduce DNS results, we perform
(i) fixed gradient and (ii) flux-driven simulations of the reduced
models with 5 different numbers of retained poloidal modes, and we
compare the results to $1024^{2}$ padded resolution DNS, using a
pseudo-spectral solver.

\begin{table}[tbph]
\centering{}%
\begin{tabular*}{1\columnwidth}{@{\extracolsep{\fill}}@{\extracolsep{\fill}}ccc}
\hline 
Padded resolution & Domain size $L_{x},L_{y}$ & Dissipation $\nu,D$\tabularnewline
$N_{px}\times N_{py}=1024^{2}$ & $20\pi/k_{y0}$ & $0.02\cdot\gamma^{+}_{max}/k^{2}_{y0}$\tabularnewline
\hline 
\end{tabular*}\caption{Parameters of the DNS, where the injection scale $k_{y0}$ and rate
$\gamma^{+}_{max}$ depend on the remaining parameters $C$ and $\kappa$.}
\label{tab:GeneralDNS}
\end{table}

The parameters of the DNS are given in Table \ref{tab:GeneralDNS}.
As in Ref. \citealp{guillon:2025}, the domain size $L_{x}\times L_{y}$
is taken to be 10 times the injection scale $2\pi/k_{y0}$, in order
to ensure similar energy injections with varying linear parameters
$C$ and $\kappa$. Similarly, the small scale dissipation is also
scaled according to the injection scale and rate $\gamma^{+}_{max}$,
in order to compensate the injection from the linear instability.
We take it as low as possible in order to minimise viscous effects,
except at the smallest scales.

\begin{table*}[t]
\centering{}%
\begin{tabular*}{1\textwidth}{@{\extracolsep{\fill}}@{\extracolsep{\fill}}lccccc}
\hline 
Poloidal resolution $n_{y}$ & 1 & 2 & 4 & 10 & 20\tabularnewline
\hline 
Selected wave-numbers $k^{j}_{y}$ for $j\in[1,n_{y}]$ & $k_{y0}$ & $jk_{y0}/2$ & $jk_{y0}/2$ & $jk_{y0}/5$ & $jk_{y0}/10$\tabularnewline
Corresponding padded resolution & $1024\times6$ & $1024\times10$ & $1024\times16$ & $1024\times34$ & $1024\times64$\tabularnewline
\hline 
\end{tabular*}\caption{Selected poloidal wave-numbers and corresponding padded resolution
for each PTM with poloidal resolution $n_{y}$.}
\label{tab:PTRMselect}
\end{table*}

\begin{figure*}[t]
\centering{}\includegraphics[width=1\textwidth]{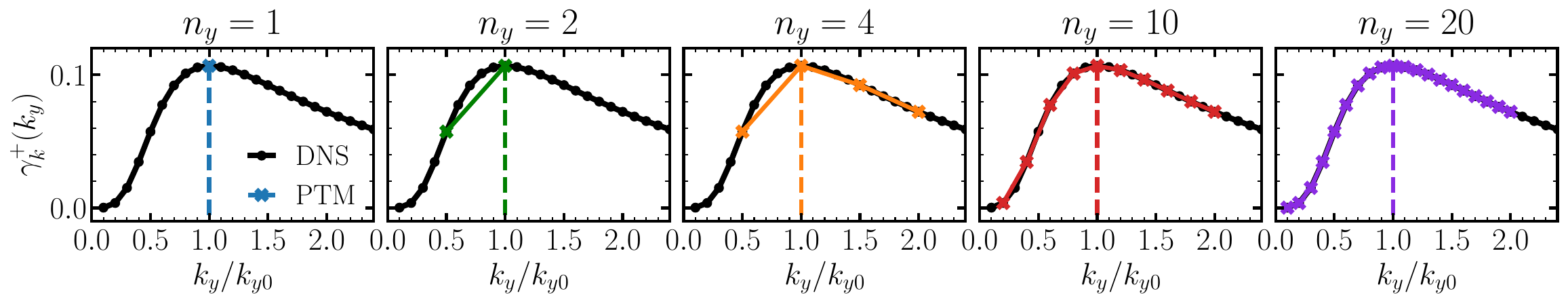}\caption{Selected wave-numbers $k^{j}_{y}$ (coloured crosses), normalised
by the most unstable wave-number $k_{y0}$, and their coverage of
the growth rate $\gamma^{+}_{k}(k_{y})$ (black line), for the 5 PTMs
with poloidal resolution $n_{y}$ given in Table \ref{tab:PTRMselect}.
The most unstable mode $k_{y0}$ is shown by the dashed line, and
the modes in the DNS grid are shown by the black dots. Coloured solid
lines show how the injection is approximated in PTMs.}
\label{fig:PTRMdistrib}
\end{figure*}

\begin{table*}[t]
\begin{tabular*}{1\textwidth}{@{\extracolsep{\fill}}lccccc}
\toprule
Model & $n_{y}=1$ & $n_{y}=2$ & $n_{y}=4$ & $n_{y}=10$ & $n_{y}=20$\tabularnewline
\hline 
\noalign{\vskip0.1cm}
\begin{cellvarwidth}[m]
\raggedright Turbulent \\
self-interactions
\end{cellvarwidth} & None & \begin{cellvarwidth}[m]
\centering
Only of the type

$\left(k_{x},k_{y0}\right)$, $\left(k^{\prime}_{x},-k_{y0}/2\right),\left(-k_{x}-k^{\prime}_{x},-k_{y0}/2\right)$
\end{cellvarwidth} & \multicolumn{3}{c}{As in a regular grid}\tabularnewline
\bottomrule
\end{tabular*}\caption{Turbulent self-interactions retained in each PTMs. All models have
ZF/turbulence interactions.}\label{tab:interact}
\end{table*}

We then construct 5 different PTMs, with $n_{y}=1$, 2, 4, 10 and
20 poloidal modes respectively. The selected poloidal wave-numbers
are listed in Table \ref{tab:PTRMselect}, along with the corresponding
padded resolution. The radial resolution $N_{px}$, box size $L_{x}$
and the exact same dissipation coefficients $\nu$ and $D$ are taken
from the DNS with the corresponding parameters $C$ and $\kappa$.
Note that in PTMs, small scale dissipation is applied \emph{only along
the radial axis}. In Fourier space, this amounts to decomposing $\nu k^{2},Dk^{2}\to\nu_{x}k^{2}_{x}+\nu_{y}k^{2}_{y},D_{x}k^{2}_{x}+D_{y}k^{2}_{y}$
and setting $\nu_{y}=D_{y}=0$. In any case, in DNS, $\nu k^{2}_{y}$
is carefully selected to be significant only at the highest wave-numbers,
so it would be negligible at large scale poloidal modes anyway. Indeed,
the results are quite similar with $\nu_{y}=\nu_{x}$, except when
using only one or two modes.

The distribution of the selected modes around the most unstable mode
$k_{y0}$ is shown in Figure \ref{fig:PTRMdistrib}. The case $n_{y}=1$
(blue) corresponds to the reduction in Tokam1D with $k_{y0}$ being
the most unstable poloidal wave-number for a given value of $C$ and
$\kappa$, and it thus keeps only interactions between ZF and turbulence.
Taking $n_{y}=2$ (red) with adding the first sub-harmonic $k_{y0}/2$
is the simplest generalisation of the 1D model, which allows for a
single kind of turbulent-turbulent triadic interactions (i.e. $\mathbf{k}=k_{y0}\mathbf{e}_{y}+k_{x}\mathbf{e}_{x}$,
$\mathbf{p}=-k_{y0}/2\hat{\mathbf{y}}+k^{\prime}_{x}\mathbf{e}_{x}$,
$\mathbf{q}=-k_{y0}/2\mathbf{e}_{y}-(k_{x}+k^{\prime}_{x})\mathbf{e}_{x}$
that satisfies $\mathbf{k}+\mathbf{p}+\mathbf{q}=0$). These lowest
resolution cases show us the limitations of the model, especially
in terms of reproducing the transition. Indeed, it has been observed
in Ref. \citealp{guillon:2025} that a reduced model with only 7 Fourier
modes, namely the most unstable mode, 2 zonal modes and the corresponding
side-bands, always yielded a ZF dominated state, regardless of $C$.
In contrast, adding the first sub-harmonic $k_{y0}/2$ enabled to
reproduce the transition, albeit at a different value of the control
parameter. Here, with the full radial resolution we seem to get a
transition even with only one poloidal mode, but also at a different
critical point and not as sharp as compared to the DNS.

The intermediate resolution $n_{y}=4$ (orange) introduces larger
wave-number modes, i.e. poloidal scales smaller than the most unstable
mode. For the larger poloidal resolutions, i.e. $n_{y}=10$ (red)
and $n_{y}=20$ (violets), we choose to use a regular grid in order to maximize triadic interactions, and to equally distribute the wave-numbers
around the most unstable mode $k_{y0}$, between $2k_{y0}/n_{y}$
and $2k_{y0}$, to have a reasonable coverage of the form of the growth rate as a function of $k_{y}$. After trial and error, we identified this form to also
be the best choice in order to reproduce the transition, as further
discussed in Section \ref{sec:fixedgrad}. Indeed, with only poloidal
scales larger than $k_{y0}$, the transition is dramatically shifted
to high $C$, and it is much harder to form ZFs, because large poloidal
scales remove energy from ZFs. Conversely, if we only include scales
smaller than $k_{y0}$, the transition is shifted towards small $C$,
and it is too easy to form ZFs because these poloidal scales feed
them. This is further justified by the analysis of the non-linear
transfers induced by each poloidal scale, as shown in Section \ref{sec:pollines}.

The retained turbulent self-interactions are summarised in Table \ref{tab:interact}.
For $n_{y}\geq4$, the complexity of the triadic interactions are
as in a DNS regular grid, although much more coarsen in the poloidal
direction. 

Note also that $n_{y}=20$ conserves all the DNS modes such that $k_{y}<2k_{y0}$,
and thus it represents an anisotropic LES without any closure term.
Conversely, the other cases correspond to different levels of reduction
of the Fourier space resolution as they increase the Fourier grid
increment $\Delta k_{y}$, which amounts to reducing the box size
in the poloidal direction since $\Delta k_{y}=2\pi/L_{y}$. Other
possible reductions that are considered while scanning $n_{y}$ and
$\Delta k_{y}$ are discussed in Appendix \ref{sec:Other-reductions},
for the fixed gradient case.

One important point is that since we choose the $k_{y0}$ to be the
most unstable mode, the selected wave-numbers depend on the linear
parameters $C$ and $\kappa$. While for the fixed gradient formulation,
this is not a problem since the parameters stay constant during the
simulation, it might be an issue for the flux-driven case (or when
changing $C/\kappa$ in a fixed gradient hysteresis scan), since the
most unstable wave-number $k_{y0}$ changes as a function of the mean
gradient $\kappa(t)$, which evolves during the simulation. In practice
however, this does not seem to cause any substantial issues in the
flux-driven simulations that we performed in this paper, either because
the poloidal resolution and coverage of the wave-number domain was
sufficient, or the change in the wave-number was not substantial.
Nevertheless, if the actual $k_{y0}$ changes dramatically, one may
have to self-consistently evolve the wave-number grid as well. This
may be done for example by stopping the simulation, modifying the
grid, interpolating the final state of the stopped simulation to the
new grid, and using that as the initial condition when restarting
the simulation.

\section{Fixed-gradient system}

\label{sec:fixedgrad}

In this Section, we apply the 5 PTMs that we previously presented,
to the fixed-gradient HW system (\ref{eq:hw}), and compare their
results with those of the DNS, scanning the adiabaticity parameter
$C$ while keeping $\kappa=1$ fixed. We use the transition from turbulence
to ZFs (which is at $C/\kappa=0.1$ in DNS) as the first criterion
of feasibility. Simulations are performed using a pseudo-spectral
method in a doubly periodic domain. The fields are initialised with
a Gaussian seed in Fourier space centred at $(0,0)$ with a low amplitude
$A=10^{-4}$ and random initial phases. Simulations are run until
$t_{f}=200/\gamma^{+}_{max}(C)$, which is roughly 10 times the time
needed to reach non-linear saturation from the initial conditions,
so that a wide time range in the turbulent saturated regime is covered.
Other settings are given in Tables \ref{tab:GeneralDNS} for DNS and
\ref{tab:PTRMselect} for PTMs.

\subsection{Evolution of the zonal flow profile}

First, we compare the spatio-temporal evolution of the zonal velocity
profile $\overline{v}_{y}(x,t)=\partial_{x}\overline{\phi}(x,t)$,
for two cases: (i) well into 2D turbulence ($C=0.01$) and (ii) ZF
dominated ($C=2$) regimes, between DNS and the 5 PTMs with different
poloidal resolutions $n_{y}$.

\begin{figure*}[tbph]
\centering{}\includegraphics[width=1\textwidth]{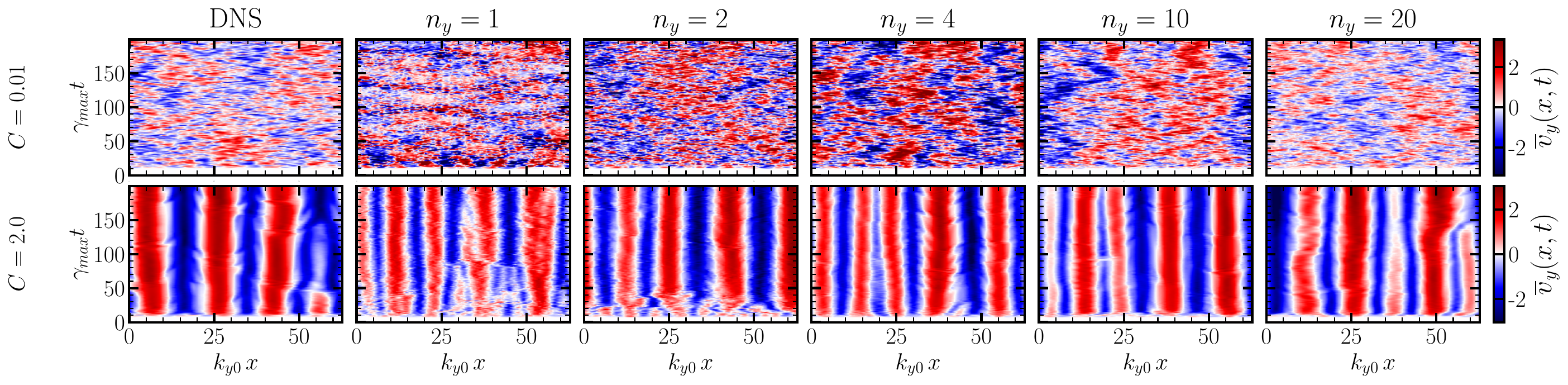}\caption{Spatio-temporal evolution of the zonal velocity profile $\overline{v}_{y}(x,t)$.
Top row: $C=0.01$ (turbulent regime), bottom row: $C=2$ (ZF regime).
DNS (leftmost column) results compared to PTMs. The radial axis ($x$
axis) is normalised by $k_{y0}(C)$, and time is normalised by $\gamma^{+}_{max}(C)$.}
\label{fig:hovmollervyfg}
\end{figure*}

The results are shown in Figure \ref{fig:hovmollervyfg} (top row:
$C=0.01$, bottom row: $C=2$). At first glance, all the models seem
to reproduce the two different regimes observed in DNS, qualitatively.
In the turbulent regime ($C=0.01)$, we observe a chaotic state without
any large scale flow, while the ZF regime ($C=2$) is dominated by
quasi-steady jets of roughly similar radial size.

However, one can see that, for the lowest poloidal resolutions $n_{y}=\left\{ 1,2\right\} $,
the zonal structures (bottom row) fluctuate, especially their edges,
and undergo several merging events. On the contrary, for larger resolutions
$n_{y}=\left\{ 4,10,20\right\} $, the zonal velocity profiles are
more steady. In the turbulent case (top row), the velocity variations
appear intermittent for $n_{y}=1$, while they are smoother for higher
$n_{y}$, as in the DNS. This may be due to fronts forming in the
$n_{y}=1$ case, where the system is almost 1D and might behave like
the Burgers equation. It may also be that in this case, the single
poloidal row is forced to carry both inverse and forward cascades,
and thus keeps oscillating between forward and backward fluxes, resulting
in an intermittent evolution of the velocity profiles.

Thus, even though the difference is not striking, and looking only
at these results they all appear to be acceptable, we can nonetheless
say that in terms of the spatio-temporal evolution of the poloidal
velocity, $n_{y}=4$ appears to be the best compromise among the cases
that we considered. This is further discussed in the following when
the transition between turbulence and ZFs and the scaling of the particle
flux are considered.

\subsection{Transition and flux}

\begin{figure}[tbph]
\centering{}\includegraphics[width=1\columnwidth]{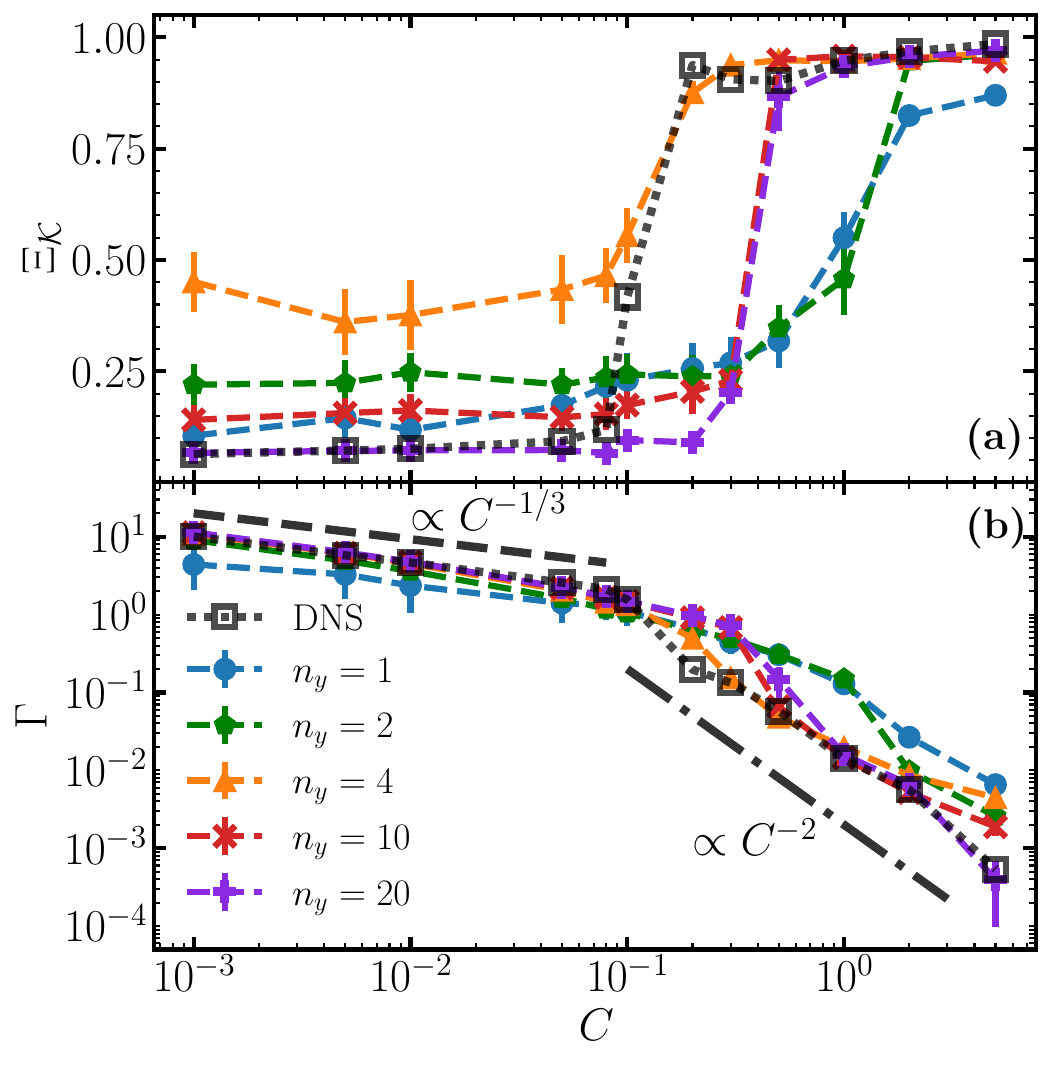}\caption{ZF level $\Xi_{\mathcal{K}}=\mathcal{K}_{ZF}/\mathcal{K}$ (a) and
mean radial particle flux $\Gamma$ (b) versus adiabaticity parameter
$C$ for DNS (black), and PTMs with $n_{y}=1$ (blue), $n_{y}=2$
(green), $n_{y}=4$ (orange), $n_{y}=10$ (red) and $n_{y}=20$ (violet).
Both quantities are averaged over $100<\gamma^{+}_{max}t<200$, with
error-bars corresponding to the standard deviation. The two power
law scalings of the particle flux are shown by the dashed ($C^{-1/3}$)
and dashed-dotted ($C^{-2}$) lines, respectively.}
\label{fig:Cscan}
\end{figure}

To assess the performance of PTMs in reproducing the transition from
turbulence to ZFs, observed in DNS at $C/\kappa=0.1$, we perform
a parameter scan over the range $C\in[10^{-3},5]$. For each simulation,
we measure the ZF level (ratio of zonal to total kinetic energy) $\Xi_{\mathcal{K}}=\mathcal{K}_{ZF}/\mathcal{K}$,
and the mean radial particle flux $\Gamma$. To fully validate a reduction,
we require a sharp transition in the ZF level at $C=0.1$, and that
the particle flux exhibit two scalings, with the transition between
the scalings corresponding roughly to the transition between the zonal
states. The results are shown in Figure \ref{fig:Cscan} for the ZF
level (top), and the particle flux (bottom), with the DNS scan in
black dots. Both quantities are averaged over $100<\gamma^{+}_{max}t<200$.

\subsubsection{Zonal transition}

All models qualitatively display a transition between 2D turbulence
(small $C/\kappa$, low ZF level $\Xi_{\mathcal{K}}$), and the quasi-1D
ZF dominated regime (high $C/\kappa$, $\Xi_{\mathcal{K}}\approx1$).

Contrary to Ref. \citealp{guillon:2025}, where a reduced system with
taking only one poloidal mode, two zonal modes and their corresponding
side-bands was always found in the zonal dominated state, here we
observe a transition even for the $n_{y}=1$ (blue) case, even though
only non-linear interactions in this model are with the zonal modes.
However, the transition in this model is much less abrupt and shifted
towards higher $C$ (i.e. $C\approx1$), compared to the DNS (see
black dots in Figure \ref{fig:Cscan} (a) and both Refs. \citealp{numata:2007,guillon:2025}).
This gradual transition is also reported in Ref. \citealp{panico:2025},
although a direct comparison remains limited, since in this previous
work, the scan was performed in the flux-driven system, with much
larger dissipation coefficients and keeping the same poloidal wave-number
while varying the control parameter $C/\kappa$.

A slightly sharper transition is observed for $n_{y}=2$ (green),
where the 1\textsuperscript{st} sub-harmonic of the most unstable
mode is also present, although it remains gradual and shifted towards
large $C$. In both cases $n_{y}=\{1,2\}$, the range over which the
zonal fraction varies is narrower than the DNS: $\Xi_{\mathcal{K}}$
is overestimated for $n_{y}=2$ in the turbulent state, or too low
for $n_{y}=1$ in the ZF state.

For the intermediary resolution $n_{y}=4$ (orange), we observe an
abrupt transition at the correct transition point $C=0.1$. However,
the zonal fraction remains overestimated in the turbulent regime (from
$35\%$ to $45$\%), which makes the existence of this state quite
sensitive to dissipation (i.e. slightly increasing the dissipation
makes the turbulent state fade away and with it the transition from
zonal flows to turbulence). This is probably due to the ratio of the
number of zonal modes to non-zonal modes being large compared to DNS
(i.e. if all the modes had equal energy it would be 1/4 vs 1/340).
In DNS, computing the zonal fraction using only the poloidal modes
that are part of the reduced model, does indeed result in a similar
overestimation of the zonal fraction in the turbulent state.

Last, for $n_{y}=10$ (red), and $n_{y}=20$ (violet), we clearly
see the sharp transition between the turbulence state with a low ZF
level, and the ZF dominated state with $\Xi_{\mathcal{K}}\approx1$.
This sharpening is expected when the resolution is increased. Indeed,
the idea of a phase transition would make sense only in the thermodynamic
limit, which theoretically requires an infinite number of modes (or
infinite resolution). While this is clearly not achieved with only
one or two poloidal modes, we are closer to it when an increasing
number of modes are included, as this dramatically increases the number
of triadic interactions.

Nevertheless, all reductions, except for $n_{y}=4$, fail to yield
the correct transition point $C/\kappa=0.1$ between the 2D and quasi-1D
regimes. While weakly resolved models overestimate the critical point
by an order of magnitude, there is still a factor 3 for $n_{y}=10$
and $n_{y}=20$, and we observed that even using $n_{y}=41$ (yet
corresponding to a padded resolution of $1024\times128$) fails to
give the correct threshold (not shown). We think that the success
of the $n_{y}=4$ is also probably coincidental. This suggests that,
increasing $n_{y}$, the convergence towards DNS is not necessarily
monotonous (but it also depends on the $k_{y}$ modes that are added).

To recover the transition point, one solution is to increase the small-scale
dissipation coefficients $\nu$ and $D$ (compared to the values used
in the original scan, which are \emph{the same} as in DNS), which
results in the critical point shifting towards smaller $C/\kappa$
(see Figure \ref{fig:Cscan-diss} (a) in Appendix \ref{sec:Diss}).
While this can be adjusted to yield the correct transition point for
$n_{y}=10$ and $n_{y}=20$, it often results in the ZF level being
overestimated in the turbulence regime for the low resolution PTMs
$n_{y}=1$, $n_{y}=2$ and $n_{y}=4$, and can make the transition
completely disappear. The shift of the critical point with increasing
viscosity and diffusion is not surprising, since dissipation is applied
at small scales and only to the non-zonal perturbations. When we increase
the diffusion coefficient, we raise the dissipation of non-zonal energy
at small scales, thus favouring the formation of ZFs at lower $C/\kappa$,
where the system was previously in the turbulent regime. This effect
is also observed in DNS \citep{guillon:2025}, although it remains
much smaller.

However, tuning small-scale dissipation is not satisfactory: it introduces
an additional parameter, which amounts to using an arbitrary closure
that requires \emph{a priori} knowledge of the fully resolved system,
and it shows strong dependence on the level of reduction. Nonetheless,
better corrections/closures require a more detailed understanding
of the physics of the transition mechanism, which is unfortunately
still lacking at that time.

\subsubsection{Particle flux}

The observations above apply equally to radial particle flux. All
PTMs reproduce, to some extent, the dual power law scaling of the
flux with $C$: a slow decrease in the turbulent regime, and then
a much steeper drop in the ZF regime, which is consistent with the
scalings that are previously observed in DNS \citep{hu:1997,guillon:2025}.

In the turbulent regime, $C<0.1$, all models reproduce the particle
flux, as evidenced by all the curves collapsing in that regime in
Figure \ref{fig:Cscan} (b), except for the most reduced model ($n_{y}=1$),
which underestimates the flux. This is probably due to ZF level being
overestimated in the turbulent regime for that model, together with
an energy injection that is too low, since the growth rate has been
severely truncated, as shown in Figure \ref{fig:PTRMdistrib}.

In the ZF dominated regime, for $C\gg0.1$, all the models display
a steeper decrease with $C$, but only high resolution PTMs, i.e,
$n_{y}=10$, $n_{y}=20$ and, to some extent, $n_{y}=4$, quantitatively
reproduce the particle flux. In contrast low $n_{y}$ models tend
to overestimate it.

As for the transition, all the models fail to predict the flux just
above the critical point $C\gtrsim0.1$, and exhibit a power law scaling
close to $C^{-1/3}$, which is the scaling observed in the turbulent
regime. This is consistent with the transition point shifting to higher
$C$ observed for the zonal energy fraction. To recover the correct
transition value where the two scalings intersect, which is roughly
the critical point $C/\kappa=0.1$, one can increase the value of
numerical dissipation, as shown in Figure \ref{fig:Cscan-diss} (b)
in Appendix \ref{sec:Diss}. However, as mentioned before, we find
this unsatisfactory, since it introduces a fitting parameter, which
we have no way of fixing without running the DNS or an actual, detailed
understanding of the transition mechanism.

\subsection{Numerical performance of the models}

In Table \ref{tab:runtimes} we compare runtimes between the DNS and
each PTM, for both $C=0.01$ and $C=1.0$, run until $t=200/\gamma^{+}_{max}$.
Simulations are run on a single NVIDIA GeForce RTX 3060 GPU.

\begin{table}[tbph]
\centering{}%
\begin{tabular*}{1\columnwidth}{@{\extracolsep{\fill}}@{\extracolsep{\fill}}lcccccc}
\hline 
$C$ & DNS & $n_{y}=1$ & $n_{y}=2$ & $n_{y}=4$ & $n_{y}=10$ & $n_{y}=20$\tabularnewline
\hline 
$1$ & 3.8 h & 13 min & 12 min & 7 min & 8 min & 7 min\tabularnewline
$0.01$ & 9.1 h & 22 min & 24 min & 34 min & 38 min & 41 min\tabularnewline
\hline 
\end{tabular*}\caption{Comparison of runtimes between DNS and each PTM with poloidal resolution
$n_{y}$, for $C=0.01$ and $C=1.0$, run until $t=200/\gamma^{+}_{max}$.}
\label{tab:runtimes}
\end{table}

PTMs run in average 25 time faster than the DNS in the ZF regime,
and 15 times faster in the turbulent regime. They are faster in the
ZF regime because the flows are stationary and turbulence is strongly
suppressed. The reason why $n_{y}=1$ and $n_{y}=2$ runs take longer
than the other models for $C=1$ is because they are closer to the
turbulent regime than the ZF regime for these models, since the transition
point is shifted to higher $C$. Note that we could optimise the padded
resolution along the poloidal direction so that it corresponds either
to a power of two, or a power of two times a small prime number, in
order to increase the computational speed. For all these reasons,
we underline that the speed-up factor should not be over-generalised.
Even though our goal here is not to perform an exhaustive performance
benchmark, we can also note that, while the DNS runs considerably
slower on CPU, the reduced models typically run at comparable speeds
on GPU and on CPU, at least for these resolutions.

\subsection{Discussion on the results}

From the fixed gradient study, two candidates stand out as good compromises
between reduction vs. fidelity. The first one is $n_{y}=4$, that
happens to have the correct transition point, but overestimates the
ZF level in the turbulence regime, because of the ratio of the number
of zonal to non-zonal modes. However, as a result of this imbalance,
the behaviour of this model is very sensitive to dissipation. The
second model, $n_{y}=10$, which has equal number of (non-zonal) modes
on both sides of $k_{y0}$, displays an abrupt transition that jumps
from a very low zonal fraction in the turbulence regime, to almost
one in the zonal regime. However, its transition point is shifted
to $C\approx0.3$.

Models with lower resolutions cannot reproduce either the transition
or the particle flux. Note that we tried intermediary resolutions
between $n_{y}=4$ and $n_{y}=10$, see Figure \ref{fig:Cscan-intermediary}
in Appendix \ref{sec:Other-reductions}, but they only slightly improve
the sharpening of the transition compared to $n_{y}=4$, and we need
to go to $n_{y}=10$ to see an actual improvement.

We also tried making the wave-number distribution around $k_{y0}$
asymmetric, i.e. adding modes mainly with $k_{y}<k_{y0}$, or with
$k_{y}>k_{y0}$, but this led, to turbulence dominated systems even
at high $C$ with most $k_{y}<k_{y0}$, and to ZFs dominated systems
even at low $C$ with most $k_{y}>k_{y0}$ (see Figure \ref{fig:Cscan-other}
in Appendix \ref{sec:Other-reductions}).

Surprisingly, going to better resolved PTMs, which are pLES models
that are truncated at different poloidal scales, does not improve
these results significantly. With $n_{y}=41$ (i.e. a $1024\times128$
padded resolution, which is already not that low, and we need to set
$\nu_{y}=\nu_{x}$ as we reach dissipative scales along $k_{y}$),
and $k^{i}_{y}\in[k_{y0}/10,4k_{y0}]$, we get an abrupt transition,
but at $C\approx0.05$. Notice that, even though the box length are
the same in both directions, there is an influence of the differences
in resolutions of the radial and the poloidal directions.

Finally, observing the hysteresis loop in the PTMs is non-trivial,
and requires choosing the correct domain size in the poloidal direction,
since the linear properties change when we dynamically change $C$
in a simulation, while the grid remains fixed. While this was not
a problem for DNS, in the PTMs it has high consequences, and can lead
to the system never doing the backward, or even the forward transition.

As a partial conclusion, we note that although $n_{y}=\{4,10,20\}$
display the non-trivial key features of the DNS, i.e. two different
transport regimes, with a correct flux prediction in each regime for
$n_{y}=\{10,20\}$, and a sharp transition from turbulence to ZFs,
they seem to miss an essential ingredient to capture the correct transition
point and the hysteresis loop. Since increasing the poloidal resolution
up to a padded resolution of $N_{py}=128$ does not improve much,
the missing ingredient may be related to the balance of mode distribution
around the most unstable mode $k_{y0}$, or to enstrophy dissipation
at small scales which could be recovered using some closure.

It could also be related to the existence of some transition scale
between the large, adiabatic, ZF scales, and the smaller 2D turbulent
scales, similar to the so-called $k_{\beta}$ scale in $\beta$-plane
turbulence \citep{vallis:1993,gurcan:2024}. If a similar scale exists
for the transition in the HW system, PTMs may need to have poloidal
modes correctly distributed on each side of this scale. In any case,
a detailed understanding of the transition is needed to design better
reduced models, but the results from PTMs give us insight into the
underlying mechanisms. The current results are nevertheless encouraging,
showing that the non-trivial behaviour of the system can be observed
with only 4, or 10 modes, and that an appropriate closure term should
be enough to recover the correct transition point, instead of increasing
the number of modes.

Here, these results are obtained for the fixed-gradient formulation.
In the following, we will test the models in a flux-driven formulation,
which is admittedly more relevant for fusion plasma applications.
Since in the latter, the mean profile or gradient, i.e. $\kappa$,
evolves as a response to sources, sinks, which impose an average turbulent
flux, it offers another channel of energy transfer/dissipation, which
may change the conclusions of the study based on the fixed-gradient
formulation.

\section{Flux-driven system}

\label{sec:fluxdriven}

We now consider the particle flux-driven HW system \citep{guillon:2025c},
where the zonal part of the density equation (\ref{eq:hw2}) is replaced
by a transport equation (\ref{eq:nrfluxdriven}) on the total radial
density profile $n_{r}$, where source and sink terms are introduced.
In this formulation, the mean density gradient $\kappa$ is no longer
imposed, but rather, evolves as a response to a turbulent particle
flux, relaxing the hypothesis of scale separation between transport
and turbulence.

In this case, in addition to yielding the same ZF level as in DNS,
we also require the candidate reduced model to converge towards the
same steady-state, and thus converge to the correct final mean density
profile. Since the stationary state corresponds to a balance between
sources, sinks and the turbulent flux, the steady-state flux should
be the same as that of the DNS. However, its statistics --- standard
deviation, large events, etc. --- might be different. So, it makes
more sense to compare the probability distribution functions (PDFs)
of the particle flux rather than its mean value. Thus, in order to
assess the performance of different reduced models, we test them in
the turbulent ($C/\kappa=0.01)$ and in the ZF regimes ($C/\kappa=1)$.
We add a particle source at the inner radial part, and we compare
the results of PTMs and of the DNS.

\subsection{Simulations set-up}

In order to perform simulations of the flux-driven system using a
pseudo-spectral method, which is more convenient since PTMs are naturally
defined in Fourier space, we use the recently-developed P-FLARE code
\citep{guillon:2025a,guillon:2025c}, which relies on the penalisation
method, allowing us to impose boundary conditions on the density profile.
In practice, we use the equivalent of a Neumann boundary condition
(free moving) at the inner radius, and a Dirichlet boundary condition
at the outer radius (clamped), while the periodicity along the poloidal
direction is maintained.

Note that since we penalise both ends of the radial axis, the size
of which is $L_{x}$, the latter is divided in two regions: the \emph{physical
domain}, $x\in[x_{b1},x_{b2}]$, at the centre of the box, and the
\emph{buffer zone}, $x\in[0,x_{b1}]\cup[x_{b2},L_{x}]$, located at
the ends, where $x_{b1}$ and $x_{b2}$ are respectively the inner
and outer boundaries between the physical domain and the buffer zones.
Boundary conditions are applied at these boundary points, and beyond
them, the fluctuations are strongly damped. Therefore, in the following,
we only consider the physical domain $x\in[x_{b1},x_{b2}]$, and compute
mean gradients between these two points. Details of the penalisation
method for each simulation is given in Appendix \ref{sec:Penalisation-parameters},
and the general methodology is presented in Ref. \citealp{guillon:2025c}.

We now consider the 4 following PTMs: $n_{y}=\{1,2,4,10\}$, with
same parameters as in Table \ref{tab:PTRMselect}. We drop $n_{y}=20$
since it behaves very similarly to $n_{y}=10$.

\subsection{Flux-driven in turbulence and zonal flow dominated cases}

\label{subsec:Flux-driven-farmargi}

In order to study the capabilities of the flux-driven PTMs, we first
perform two DNS of padded resolution $1024^{2}$ with $C=0.01$ and
$C=1$ and the following initial density profile: 
\begin{equation}
n_{r}(x,t=0)=L_{x}\exp\left[-4\left(x/L_{x}\right)^{2}\right],\label{eq:nr0}
\end{equation}
where $L_{x}$ is the size of the total radial domain (i.e. physical
and buffer zones). This initial profile yields an initial mean gradient
$\kappa\approx1.2$, so that each simulation is initially well into
the turbulent ($C/\kappa(t=0)\approx0.01$), and the ZF dominated
regimes ($C/\kappa(t=0)\approx1$), respectively. The domain size
(hence the Fourier grid) and dissipation are set exactly as for the
fixed gradient case, see Table \ref{tab:GeneralDNS}. Note that, since
we first expect a relaxation of the initial profile following the
saturation of the linear instability, we prefer to use the most unstable
mode $k_{y0}$ obtained with $\kappa=1$, rather than with the initial
mean gradient $\kappa\approx1.2$ obtained from Eq. \ref{eq:nr0}.
Indeed after this relaxation, the resulting non-linear state will
have a slightly lower $\kappa$. Penalisation parameters are given
in Appendix \ref{sec:Penalisation-parameters}.

\begin{table}[tbph]
\centering{}%
\begin{tabular*}{1\columnwidth}{@{\extracolsep{\fill}}@{\extracolsep{\fill}}cccc}
\hline 
$C$ & $\alpha$ & $x_{0}$ & $\sigma_{n}$\tabularnewline
\hline 
0.01 & 1 & 135$\cdot L_{x}/N_{x}$ & $0.02\cdot L_{x}$\tabularnewline
1 & 0.1 & 135$\cdot L_{x}/N_{x}$ & $0.02\cdot L_{x}$\tabularnewline
\hline 
\end{tabular*}\caption{Source parameters for the two flux-driven simulations with $C=0.01$
and $C=1$. Note that the source centre $x_{0}$ is proportional to
the radial grid increment $L_{x}/N_{x}$, where $N_{x}=2\lfloor N_{px}/3\rfloor$
is the unpadded resolution, which changes with $C$, by definition
of $L_{x}$ in Table \ref{tab:GeneralDNS}.}
\label{tab:FDDNS1}
\end{table}

We also introduce a particle source $S(x)$ of constant amplitude
$\alpha$, centred close to the inner boundary $x_{b1}$, with a Gaussian
shape:
\[
S(x)=\frac{\alpha}{\sigma_{n}\sqrt{2\pi}}\exp\left[-\left(x-x_{0}\right)^{2}/\left(2\sigma^{2}_{n}\right)\right],
\]
where $x_{0}$ is the centre of the source, and $\sigma_{n}$ is its
width. The values for the source parameters are given in Table \ref{sec:app-eigenvalues}.
Notice that $\alpha=0.1$ for $C=1.0$, while $\alpha=1$ for $C=0.01$,
in accordance with the level of transport that is expected for the
same gradient in each case.

The initial zonal velocity profile is set to zero, and non-zonal fluctuations
are initialised with the same seed method as for the fixed gradient
case. The DNS are run until $t_{f}=300/\gamma^{+}_{max}$ for $C=0.01$
and $t_{f}=200/\gamma^{+}_{max}$ for $C=1$, so that we reach a well
established saturated state. We then use these final states as initial
conditions for the 4 PTMs, where we project the DNS Fourier fields
on the reduced grids (see Figure \ref{fig:PTRMdistrib}). Finally,
we run both the DNS and the PTMs for another $200/\gamma^{+}_{max}$
and compare the results for this time interval. We choose to initialise
the PTMs with the saturated state of the DNS in order to remove the
initial profile relaxation that happens after the linear instability
saturates. Since the initial phase may involve larger scale poloidal
wave-numbers, because the initial profile is steeper, this choice
is more favourable for the PTMs. In fact, the more stringent test
of starting PTM runs from the initial Gaussian profile (\ref{eq:nr0}),
instead of projecting the DNS turbulent state, does not change the
results for $n_{y}=4$ and $n_{y}=10$ much. However, for $n_{y}=1$
and $n_{y}=2$, the final states are even farther from that of the
DNS. A quantitative comparison between DNS and PTMs can also be found
in Appendix \ref{sec:quant}.

\subsubsection{The turbulent case}

\begin{figure*}[t]
\centering{}\includegraphics[width=1\textwidth]{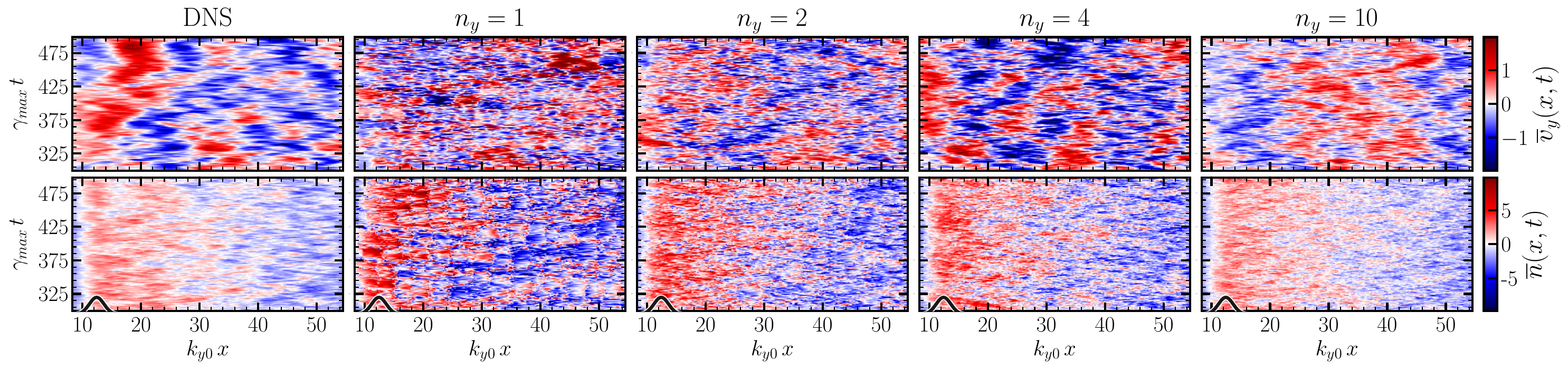}\caption{Case $C=0.01$ (turbulent regime). Spatio-temporal evolution of the
zonal velocity $\overline{v}_{y}$ (top row) and zonal density $\overline{n}$
(bottom row) profiles, for the DNS (left column) and the 4 PTMs. The
particle source is shown by the black line on the density profiles,
in arbitrary units. The radial coordinate $x$ is normalised by $k_{y0}$,
and time is normalised by $\gamma^{+}_{max}$.}
\label{fig:FDC001prof}
\end{figure*}

\begin{figure*}[tbph]
\centering{}\includegraphics[width=1\textwidth]{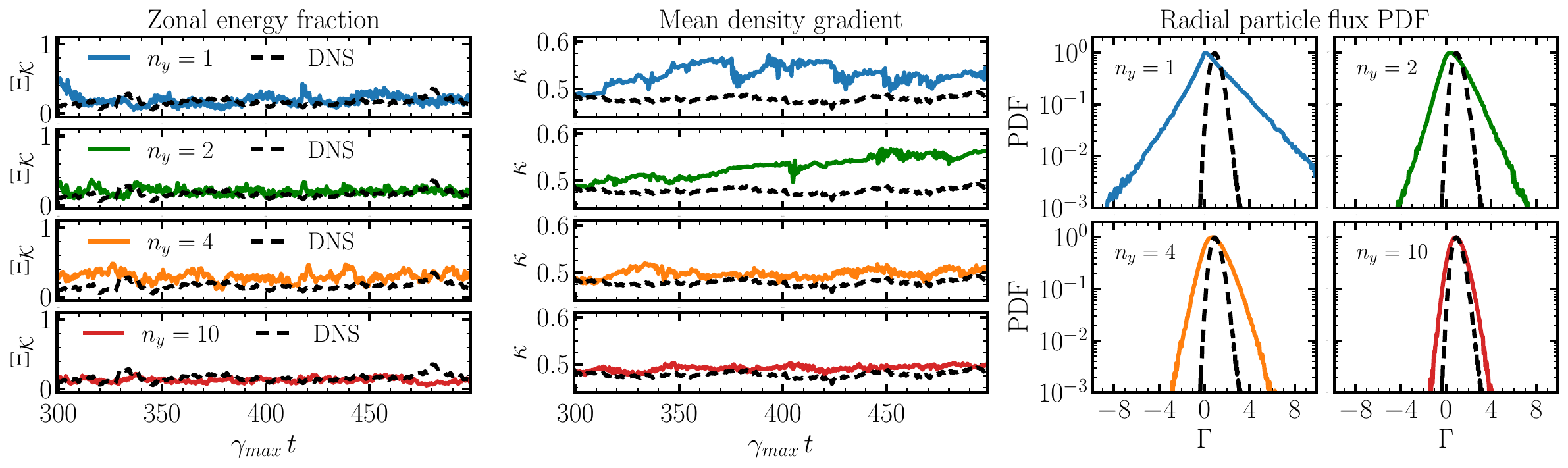}\caption{Case $C=0.01$ (turbulent regime). Time evolution of the ZF level
$\Xi_{\mathcal{K}}$ (left column) and of the mean gradient $\kappa$
(centre column) for the 4 PTMs $n_{y}=1$ (top row, blue), $n_{y}=2$
(2\protect\protect\textsuperscript{nd} row, green), $n_{y}=4$ (3\protect\protect\textsuperscript{rd}
row, orange) and $n_{y}=10$ (bottom row, red), compared to DNS (black
dashed line). Right column: PDF of the radial particle flux $\Gamma$
computed over each radial position $[x_{0},x_{b2}]$ and time, for
each PTM (coloured solid line) compared to that of DNS (black dashed
line). PDFs are normalised by their maximum value. Time is normalised
by $\gamma^{+}_{max}$.}
\label{fig:FDC001quant}
\end{figure*}

First, we consider the turbulence dominated case $C=0.01$ in Figure
\ref{fig:FDC001prof}, where we show the spatio-temporal evolution
of the zonal velocity $\overline{v}_{y}$ (top row) and zonal density
$\overline{n}$ (bottom row) profiles, for the DNS and PTMs. As for
the fixed gradient case, all the reduced models as well as the DNS
display chaotic behaviour. And as previously, the most strongly reduced
models $n_{y}=1$ and $n_{y}=2$ vary more intermittently in time,
displaying spatio-temporal structures resembling discontinuous fronts.

In Figure \ref{fig:FDC001quant}, we show the time evolutions of the
ZF level $\Xi_{\mathcal{K}}$ (left column) and the mean gradient
$\kappa$ (centre column) for the 4 PTMs, and compare the results
to the DNS shown in black dashed line. Again, we observe that $n_{y}=4$
(orange) and $n_{y}=10$ (red) perform much better than the lowest
PTMs $n_{y}=1$ (blue) and $n_{y}=2$ (green). Although the latter
models reproduce correctly the ZF level, they overestimate the mean
gradient, with strong relaxation events in time evolution, which might
correspond to the fronts observed in Figure \ref{fig:FDC001prof}.
On the other hand, higher resolution PTMs (orange and red), correctly
predict the ZF level (it is slightly overestimated for $n_{y}=4$
as for the fixed-gradient case), and their mean gradients are rather
close to the DNS results (part of the time evolution is obviously
random between the simulations).

We then compute the particle flux PDF for $300<\gamma^{+}_{max}t<500$
and $x\in[x_{0},x_{b2}]$ (we only consider the region that is to
the right of the source position $x_{0}$, since we are interested
in the outward flux, from the source to the edge). The PDF obtained
for each PTM is shown in each panel of the right column of Figure
\ref{fig:FDC001quant} (coloured solid line), and compared to that
of the DNS (black dashed line). First, considering the DNS, we see
that the PDF is skewed and exhibits deviations from a normal law for
large flux values. In earlier works \citep{basu:2003}, it has been
argued that this asymmetry corresponded to a log-normal distribution.
This distribution comes from a poloidal average of the point-wise
radial flux PDF, which itself is non-Gaussian but can be obtained
by assuming that density and velocity fluctuations are correlated
Gaussians \citep{carreras:1996a}. Note that the Gamma distribution
is also used, but rather for the distribution of the fluctuations
amplitude, instead of the flux itself \cite{garcia:2012}. However,
we found that the PDF could also be reproduced using an exponentially
modified Gaussian (EMG) law \citep{:m,dyson:1998a}, which is detailed
in Appendix \ref{sec:EMG}, and that corresponds to a random variable
which is the sum of a normal law and an exponential law. Our interpretation
is that the normal component would represent bulk turbulence and the
exponential would represent large flux events, induced by coherent
structures, avalanches, blobs etc. \citep{naulin:1999}.

Comparing DNS with PTMs, we find that $n_{y}=4$ and $n_{y}=10$ display
similar asymmetric distributions, although $n_{y}=4$ has a larger
standard deviation. Normalising PDFs by their standard deviation make
them collapse to that of the DNS (not shown here). The model $n_{y}=2$
has a shape similar to that of $n_{y}=4$, but an even larger standard
deviation. For $n_{y}=1$, however, the shape is very different, with
a sharp peak around which the PDF decreases following an exponential
distribution. It has some similarities with the point wise radial
flux PDF found in Refs. \citealp{carreras:1996a,basu:2003} (although
in these works the exponential distribution is multiplied by a Bessel
function of the flux), which might be interpreted as the absence of
ensemble averaging along the poloidal direction for $n_{y}=1$.

Thus, for the flux-driven case, the statistical properties of the
DNS --- the ZF level, the mean gradient and, to some extent, the
PDF of the radial particle flux --- are reasonably reproduced by
the $n_{y}=10$ PTM, although the PDF is slightly flatter. The model
with $n_{y}=4$ also performs decently: although it yields a larger
standard deviation for the particle flux distribution, and its ZF
level is higher (again due to the large ratio of the number of zonal
to total modes), it still gives the correct mean density gradient.
The models $n_{y}=1$, 2, however, fail to reproduce the DNS mean
gradient, and completely overestimate the fluctuations of the particle
flux.

\subsubsection{The zonal flow dominated case}

\begin{figure*}[tp]
\centering{}\includegraphics[width=1\textwidth]{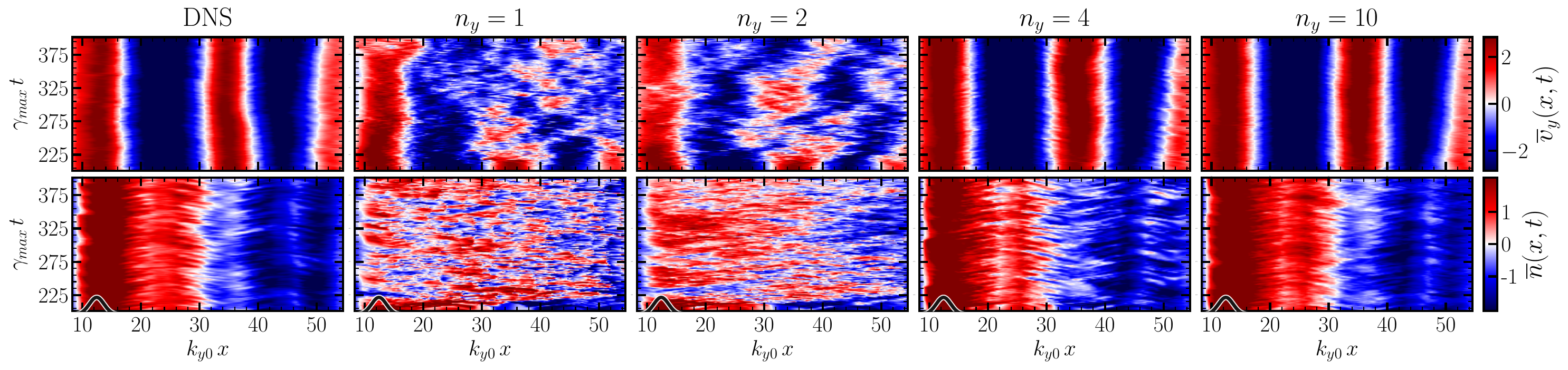}\caption{Case $C=1$ (ZF regime). Spatiotemporal evolution of the zonal velocity
$\overline{v}_{y}$ (top row) and zonal density $\overline{n}$ (bottom
row) profiles, for the DNS (left column) and the 4 PTMs.}
\label{fig:FDC1prof}
\end{figure*}

\begin{figure*}[tp]
\centering{}\includegraphics[width=1\textwidth]{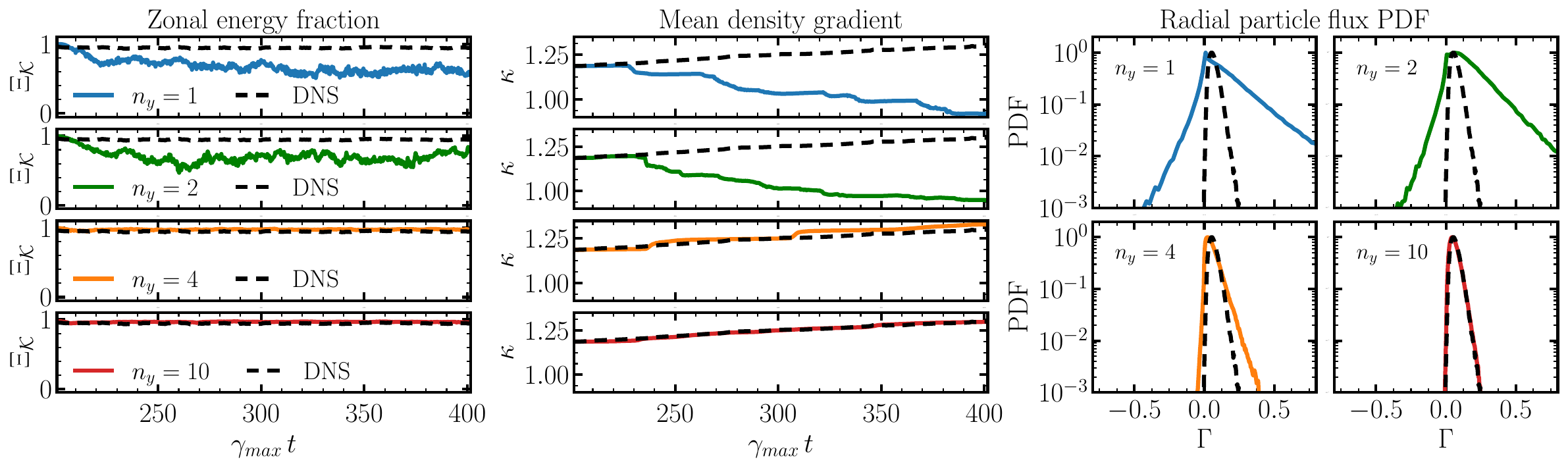}\caption{Case $C=1$ (ZF regime). Time evolution of the ZF level $\Xi_{\mathcal{K}}$
(left column) and of the mean gradient $\kappa$ (centre column) for
the 4 PTMs $n_{y}=1$ (top row, blue), $n_{y}=2$ (2\protect\protect\textsuperscript{nd}
row, green), $n_{y}=4$ (3\protect\protect\textsuperscript{rd} row,
orange) and $n_{y}=10$ (bottom row, red), compared to DNS (black
dashed line). Right column: PDF of the radial particle flux $\Gamma$
computed over each radial position $[x_{0},x_{b2}]$ and time, for
each PTM (coloured solid line) compared to that of DNS (black dashed
line).}
\label{fig:FDC1quant}
\end{figure*}

The same comparison can be done in the ZF dominated case, choosing
$C=1$. In Figure \ref{fig:FDC1prof}, we show the spatio-temporal
evolution of the zonal velocity $\overline{v}_{y}$ (top row) and
zonal density $\overline{n}$ (bottom row) profiles, for the DNS and
PTMs. Contrary to the previous turbulent case, we see here that $n_{y}=1$
and $n_{y}=2$ are not able to reproduce the ZF state, as depicted
by the rapid, early breaking of the zonal jets, and the chaotic zonal
density profile, at around $\gamma^{+}_{max}t\approx225$. In contrast,
for $n_{y}=4$ and $n_{y}=10$, the jets are maintained, and even
feature the narrowing of the rightmost jet also seen in the DNS. The
density profile is also well reproduced, with the oblique lines demonstrating
the particle transport from the source at the left towards the right
edge as well as to the troughs of the zonal velocity profile.

The inability of $n_{y}=1$ (blue) and $n_{y}=2$ (green) PTMs to
reproduce the zonal state is further illustrated in Figure \ref{fig:FDC1quant},
where the ZF level and the mean density gradient are shown as functions
of time. Low resolution PTMs underestimate the zonal fraction, which
is linked to breaking apart of the zonal jets observed in Figure \ref{fig:FDC1prof}.
Consequently the relaxation is more pronounced, and the mean gradient
decreases over time (while it increases for the DNS). On the contrary,
$n_{y}=4$ (orange) and $n_{y}=10$ (red) reduced models follow exactly
the ZF level and the mean gradient measured in the DNS.

We then consider the PDFs of particle flux for the PTMs and the DNS
in the right column of Figure \ref{fig:FDC1quant}. Again, the DNS
distribution is skewed, with deviation of large positive flux events
from a Gaussian distribution, although the flux values are 10 times
lower than in the turbulent case. For the ZF dominated case, the description
by the EMG law is less accurate (see Appendix \ref{sec:EMG}), probably
because of the transport barriers induced by ZFs, which confine turbulence
inside the jets, except for some very rare coherent structures.

Turning to PTMs, while $n_{y}=10$ matches the DNS distribution almost
exactly, $n_{y}=4$ yields a slightly larger standard deviation, with
a flatter tail for positive flux values. Thus, in this case, for $n_{y}=4$,
the shape of the distribution does not match that of the DNS. For
$n_{y}=2$, the difference is much more important, and the flux PDF
has a larger standard deviation and an even flatter tail. This is
expected since $n_{y}=2$ is not in the ZF dominated regime anyway,
due the collapse of the zonal jets observed for this model. Finally,
$n_{y}=1$ exhibits two exponential slopes as in the turbulent case,
and much larger flux events than the DNS, which is again consistent
with the model not being in the ZF dominated regime.

\subsubsection{Discussion on the results}

Overall, $n_{y}=4$ and $n_{y}=10$ PTMs reproduce the mean behaviour
of flux-driven DNS simulations far from the marginal threshold reasonably
well, both for the turbulent case and the ZF dominated case, in terms
of ZF level and mean density gradient. Nevertheless, only $n_{y}=10$
approaches the PDF of particle flux, while $n_{y}=4$ displays a larger
standard deviation, especially for the turbulent case.

On the other hand, $n_{y}=1$ and $n_{y}=2$ PTMs also reproduce,
to some extent, the turbulent state, even though the mean gradient
is overestimated and the standard deviations of the PDFs are too large,
with a qualitatively different distribution with respect to the DNS.
Nevertheless, both models completely fail to maintain the zonal structure
in the ZF dominated case, which leads to a completely different final
state. Note that however, increasing small-scale dissipation, would
result in the inverse situation where these low resolution models
would exhibit a ZF dominated state even at low $C/\kappa$.

One limitation of PTMs in the flux-driven framework might be that
the most unstable mode changes over time, which would require to re-scale
the grid at some point. But in the turbulent case, even though the
most unstable mode changed by  $25\%$, this did not seem to cause
a particular problem, given the good performance of $n_{y}=10$, where
$4k_{y0}/5$ or $6k_{y0}/5$ is also present, and can play the role
of the most unstable mode with the given profile. This, actually,
is probably one of the reasons why, while $4$ modes seemed sufficient
in the fixed gradient case, $10$ modes were clearly better for the
flux driven case.

One remaining question is the behaviour of PTMs close to the marginal
threshold $C/\kappa=0.1$ between turbulent and ZF dominated regimes.
Close to marginality, flux-driven systems can exhibit behaviour reminiscent
of self-organised criticality \citep{carreras:1996}. Such behaviour
have been observed at the threshold $C/\kappa=0.1$ in the flux-driven
HW system \citep{guillon:2025c,diamond:2025}, which we also observed
with the $n_{y}=10$ PTM (not shown here), but at its corresponding
threshold $C/\kappa=0.3$ obtained from the fixed-gradient scan in
Figure \ref{fig:Cscan}.

It has also been argued that the hysteresis loop observed in the fixed-gradient
system caused profile stiffness \citep{wolf:2003,garbet:2004,mantica:2009}
in the flux-driven system \citep{guillon:2025c}, but our initial
attempts with PTMs were not successful. Since these models do not
exhibit the hysteresis loop around the transition point in the fixed
gradient system up to moderate resolutions, we also expect profile
stiffness not to be reproduced in the flux-driven case. But once we
are able to capture the hysteresis, which might be recovered by designing
a closure term potentially including the effects of ZFs, it will naturally
also be present in the flux-driven formulation.

\section{Role of poloidal scales}

\label{sec:pollines}

In this last section, we investigate, in the fixed-gradient case,
the role played by the different poloidal modes (or scales) that are
included in PTMs. Note that each poloidal mode, denoted by an index
$i$, that is included in a PTM $(0,k^{i}_{y}=i\Delta k_{y})$, comes
with all the radial side-bands $(k_{x}\neq0,k^{i}_{y}=i\Delta k_{y})$.
Since the reduction that we discuss in this paper is based on reducing
the number of poloidal scales, we look at the role played by each
of those scales, by considering the direction of the energy and enstrophy
transfers due to each poloidal scale, and the interaction of each
scale with ZFs.

\subsection{Energy transfer per poloidal scale}

To elucidate the transfers per each poloidal scale defined by the
index $i$ using $k^{i}_{y}=i\Delta k_{y}$, we consider the turbulent
regime $C=0.01$ and try to find out if it displays forward or inverse
cascades in the radial wave-numbers, i.e. if the energy (or enstrophy)
is transferred \emph{to} smaller radial scales, or rather if the large
radial scales \emph{gain }energy (or enstrophy) from the small scales.

\begin{figure}[tbph]
\centering{}\includegraphics[width=1\columnwidth]{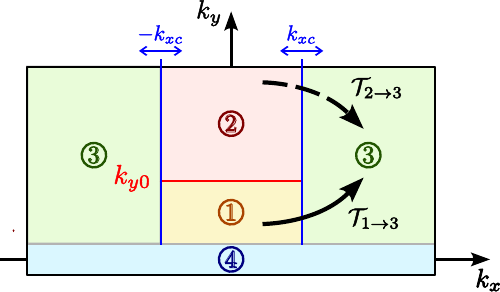}\caption{Partition of the (top-half) Fourier space into the 4 regions. The
cutoff wave-number $k_{xc}$ is varied to study the direction of the
transfers from large (1) and smaller poloidal scales (2) to small
isotropic turbulent scales (3).}
\label{fig:CLUSTER}
\end{figure}

\begin{figure*}[tp]
\centering{}\includegraphics[width=1\textwidth]{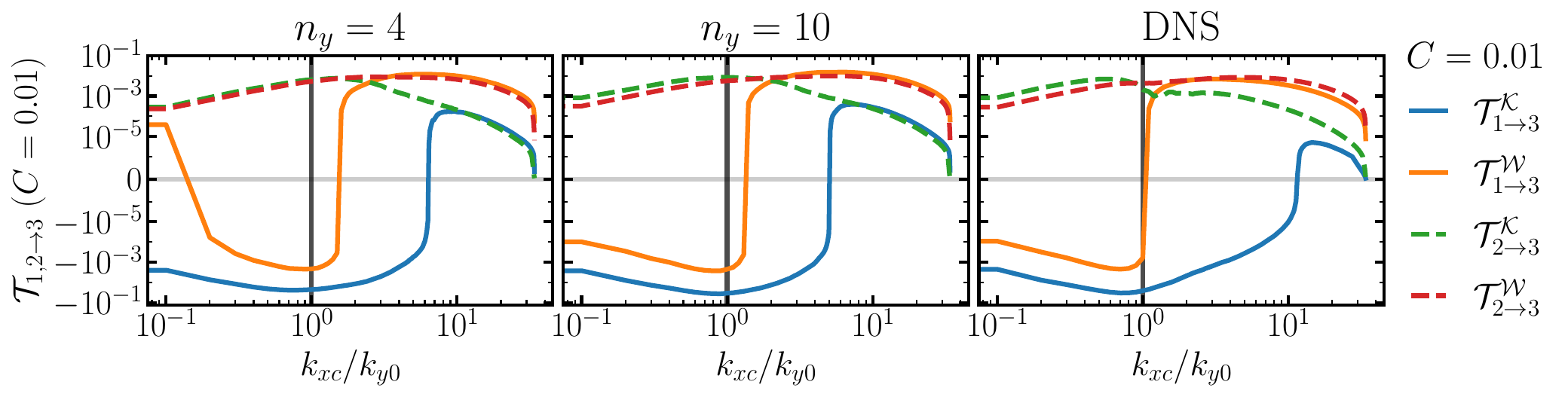}\caption{Energy $\mathcal{K}$ and enstrophy $\mathcal{W}$ transfers $\mathcal{T}^{\mathcal{K},\mathcal{W}}_{1\to3}$
(resp. $\mathcal{T}^{\mathcal{K},\mathcal{W}}_{2\to3}$) from large
(resp. small) poloidal scales to small scales, are shown in solid
blue and orange (resp. dashed green and red) lines, and averaged over
$200<\gamma^{+}_{max}t<250$, as a function of the cutoff radial wave-number
$k_{xc}$ (normalised by $k_{y0}$, shown by the vertical line). All
cases are with $C=0.01$.}
\label{fig:Tclusters}
\end{figure*}

In order to elucidate the directions of transfers we partition the
Fourier space in 4 regions, instead of focusing on single poloidal
scales. These regions, shown in Figure \ref{fig:CLUSTER}, are:
\begin{enumerate}
\item large scale, non-zonal modes, i.e. $|k_{x}|<k_{xc}$ and $0<|k_{y}|\leq k_{y0}$,
where $k_{xc}$ is a cutoff radial wave-number (region n°1, in orange);
\item small poloidal large radial scales, i.e. $|k_{x}|<k_{xc}$ and $|k_{y}|>k_{y0}$
(region n°2, in red);
\item small isotropic turbulent scales, $|k_{x}|\geq k_{xc}$ (n°3, in green);
\item zonal modes, i.e. $k_{x}$ with $k_{y}=0$ (n°4, in blue).
\end{enumerate}
This partition allows us to compute energy and enstrophy transfers
from large radial scales (i.e. $k_{x}<k_{xc}$) to small scales (i.e.
$k_{x}>k_{xc}$) separately for poloidal scales that are either larger
or smaller than the scale corresponding to the most unstable mode.
We can do so by computing the energy $\mathcal{K}$ and enstrophy
$\mathcal{W}$ transfers from the large (resp. small) poloidal scales,
i.e. region n°1 (resp. n°2), to small isotropic turbulent scales (region
n°3), which we write $\mathcal{T}^{\mathcal{K},\mathcal{W}}_{1\to3}$
(resp. $\mathcal{T}^{\mathcal{K},\mathcal{W}}_{2\to3}$). Moreover,
by varying the cutoff wave-number $k_{xc}$, we can obtain quantities
that describe radial energy transfer, very similar to an energy flux,
from large scales to small scales, that is commonly computed in isotropic
turbulence \citep[Ch. 6]{lesieur:2008}. To obtain the energy transfers,
we use a method of \emph{clustering}, detailed in Appendix \ref{sec:CLUST}.

The idea is to decompose the transfers $\mathcal{T}_{1,2\to3}$ into
a sum of transfers \emph{mediated} by each region, i.e. 
\begin{equation}
\mathcal{T}_{1,2\to3}\equiv\sum^{4}_{i=1}\mathcal{T}^{i}_{1,2\to3},\label{eq:Tclustdecomp}
\end{equation}
where $\mathcal{T}^{i}_{1,2\to3}$ is the energy (or enstrophy) transfer
from 1 (or 2) to 3, \emph{mediated} by the region n°i, which corresponds
to summing transfers due to all triads with a mode in 1 (or 2), a
mode in 3, and a mode in $i$ (see Appendix \ref{sec:CLUST}).

The results from this analysis are shown in Figure \ref{fig:Tclusters}
for $n_{y}=4$ (left), $n_{y}=10$ (middle) PTMs and DNS (right),
from the fixed gradient simulations presented in Section \ref{sec:fixedgrad},
extended to $t=250/\gamma^{+}_{max}$. The transfers $\mathcal{T}^{\mathcal{K},\mathcal{W}}_{1\to3}$
(resp. $\mathcal{T}^{\mathcal{K},\mathcal{W}}_{2\to3}$) of energy
$\mathcal{K}$ and enstrophy $\mathcal{W}$, from large (resp. small)
poloidal scales to isotropic small scales, are shown in solid (resp.
dashed) lines, as a function of the cutoff radial wave-number $k_{xc}$,
and averaged over $200<\gamma^{+}_{max}t<250$. One can see that the
large poloidal scales $k_{y}\leq k_{y0}$ receive energy ($\mathcal{T}^{\mathcal{K}}_{1\to3}<0$,
in blue) and enstrophy ($\mathcal{T}^{\mathcal{W}}_{1\to3}<0$, in
orange) from small scale turbulence, through an inverse energy cascade,
which is consistent with the 2D Navier Stokes behaviour of vorticity.
The main difference is that the enstrophy transfer is not zero in
the range of the inverse energy transfer, which may be due to injection
by a linear instability, instead of a localised random forcing. This
range extends until $k_{xc}\approx10k_{y0}$ for energy, and only
to $k_{xc}\approx k_{y0}$ for enstrophy, at which point the transfers
change sign, meaning we reach small enough scales at which turbulence
becomes isotropic. On the contrary, smaller poloidal scales $k_{y}>k_{y0}$
are always \emph{giving} their energy and enstrophy to small scale
turbulence ($\mathcal{T}^{\mathcal{K},\mathcal{W}}_{2\to3}>0)$, akin
to a forward cascade. At small radial scales, the enstrophy forward
transfer (red) is much larger than that of energy (green), hence,
we thus argue that it is the enstrophy that undergoes a forward cascade,
as in 2D Navier-Stokes turbulence.

Note that we arbitrarily choose to separate large and small poloidal
scales at the most unstable mode $k_{y0}$. However, using a cutoff
wave-number $k_{yc}<k_{y0}$, we also observed inverse transfers for
$k_{y}>k_{yc}$. Hence $k_{y0}$ appears as a natural boundary between
inverse and forward transfers, which might explain why PTMs perform
better when modes are distributed evenly around that scale. The ZF
dominated case (not shown here) is more difficult to interpret, due
to the complicated interaction with ZFs, but the main observation
is still a dual cascade, with an inverse energy transfer and a forward
enstrophy cascade. The energy cascade, however, is anisotropic, as
discussed in the following.

\subsection{Interaction with zonal flows}

To understand how different poloidal scales interact with ZFs, we
now focus on the case $C=1.0$. The time evolution of the mean zonal
kinetic energy $\mathcal{K}_{ZF}\equiv(1/2)\langle\overline{v}^{2}_{y}\rangle_{x}$
is $d_{t}\mathcal{K}_{ZF}=P_{Re}$, where $P_{Re}=-\langle\overline{v}_{y}\Pi_{\Omega}\rangle_{x}$
is the so-called Reynolds power, with $\Pi_{\Omega}(x)=-\langle\widetilde{\Omega}\partial_{y}\widetilde{\phi}\rangle_{y}$
the vorticity flux. Using Parseval's theorem on $\Pi_{\Omega}$, we
can decompose $P_{Re}=\sum_{k_{y}}P_{Re}(k_{y})$ over poloidal scales,
where 
\begin{equation}
P_{Re}(k_{y})=\left\langle \overline{v}_{y}(x)\text{Im}\left(k_{y}\widetilde{\Omega}_{k_{y}}(x)\widetilde{\phi}^{*}_{k_{y}}(x)\right)\right\rangle _{x},\label{eq:Prekydef}
\end{equation}
and $\widetilde{A}_{k_{y}}$ are Fourier transforms of $\widetilde{A}$
along $k_{y}$ (but still functions of $x$). We can thus write the
zonal kinetic energy as a sum of energy contribution from each poloidal
scale: 
\begin{equation}
\mathcal{K}_{ZF}(t)-\mathcal{K}_{ZF}(t_{0})=\sum_{k_{y}}\overline{e}_{k_{y}}(t),\label{eq:sumeky}
\end{equation}
where 
\begin{equation}
\overline{e}_{k_{y}}(t,t_{0})=\int^{t}_{t_{0}}P_{Re}(k_{y},t')\,dt'.\label{eq:eky}
\end{equation}
The reason to look at the time integral is that $P_{Re}(k_{y},t)$
itself is mostly fluctuating in the stationary state, while $\overline{e}_{k_{y}}(t,t_{0})$
keeps track of the cumulative transfer from the mode $k_{y}$ to ZFs,
starting at $t_{0}$. In practice we choose $t_{0}=200/\gamma^{+}_{max}$,
long after saturation.

\begin{figure}[tbph]
\centering{}\includegraphics[width=1\columnwidth]{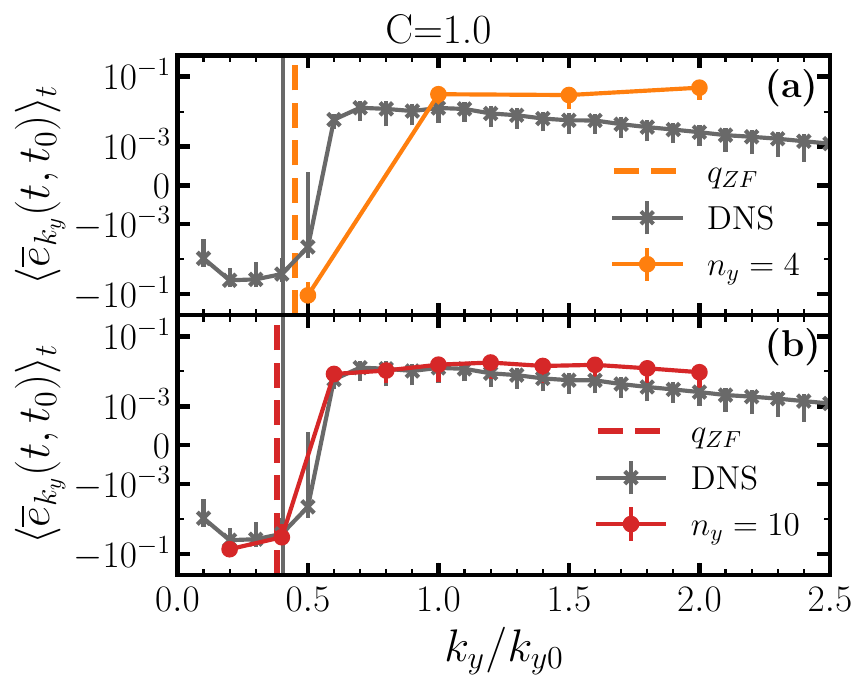}\caption{Time average of the energy $\overline{e}_{k_{y}}$ transferred from
the poloidal line $k_{y}$ to ZFs in the case $C=1$, for $n_{y}=4$
(a) and $n_{y}=10$ (b), in colored lines, both compared to DNS (grey
solid line and crosses). Averages are computed over $200<\gamma^{+}_{max}t<250$.
Vertical lines indicate the dominant zonal wave-number $q_{ZF}$ for
PTMs (dashed coloured) and DNS (solid grey).}
\label{fig:Preky}
\end{figure}

In Figure \ref{fig:Preky}, we show the energy $\overline{e}_{k_{y}}$
transferred from the poloidal scale $k_{y}$ to the zonal kinetic
energy, for the $n_{y}=4$ (a) and $n_{y}=10$ (b) PTMs (coloured
dots), compared with DNS (grey crosses), all for the case $C=1$ (using
the same simulations as in Section \ref{sec:fixedgrad}, extended
to $t=250/\gamma^{+}_{max}$). The contributions are computed and
then averaged over $200<\gamma^{+}_{max}t<250$. Note that the poloidal
wave-number is normalised by $k_{y0}$. The vertical lines indicate
the dominant zonal wave-number $q_{ZF}$ (computed as in Refs. \citealp{numata:2007}
and \citealp{guillon:2025}) for PTMs (dashed coloured) and DNS (solid
grey).

First, it is interesting to note that the energy transfer to ZFs for
the DNS changes sign with $k_{y}$. For large scale poloidal modes
$k_{y}<k_{y0}/2$, we have $\overline{e}_{k_{y}}<0$, meaning that
these scales remove energy from the ZFs. On the contrary, smaller
scales $k_{y}>k_{y0}/2$ give their energy to ZFs. Hence, there is
anisotropic inverse energy transfer from the most unstable mode to
large-scale poloidal modes, \emph{through} ZFs, in the spirit of the
inverse cascade.

Note that overall, the net Reynolds power is zero in average (see
Ref. \textcolor{magenta}{\citealp{guillon:2025d}}). Hence, the final
stationary state dominated by ZFs corresponds to a balance between
small poloidal scales feeding the ZFs (including the most unstable
mode), and the larger scales removing this energy. It is also worth
noticing that the dominant ZF scale $q_{ZF}$ is of the order of the
large poloidal scales that remove energy from the flows. Note that
in the Hasegawa-Wakatani system, one does not need to remove the energy
going to large scales (using friction or hypo-viscosity, i.e. of the
form $-1/k^{2p}$ with $p$ an integer, as commonly done in two-dimensional
turbulence studies), since the linearly damped mode can eventually
remove the energy from the system, if the phases between density and
electrostatic potential is arranged in a particular way.

PTMs $n_{y}=4$ and $n_{y}=10$ also feature a similar sign inversion
of $\overline{e}_{k_{y}}$. Interestingly, the ZF size measured for
$n_{y}=4$ is lower than any existing poloidal mode, contrary to $n_{y}=10$.
We can argue that, to correctly reproduce ZF formation and the transition,
a reduced model should at least include the poloidal scales which
allow the change in sign of $\overline{e}_{k_{y}}$, in order to have
the balance between the energy transfer into the zonal flows from
small scales and out to the large scale poloidal modes, which is inherently
impossible for $n_{y}=1$. In DNS, this balance might be broken for
$C/\kappa<0.1$, leading to the turbulent regime. To correctly reproduce
the transition, a reduced model might also need the ability to display
this imbalance in the turbulent regime. Contributions of different
poloidal scales in the formation and the saturation of zonal flows
should be studied in more detail and for other similar systems in
the future, for a better understanding of the mechanisms at play.

\section{Conclusion}

We have introduced a reduction scheme, using an anisotropic truncation,
where only a small number of poloidal modes are kept together with
all the radial wave-numbers of an initially isotropic high resolution
Fourier domain. We have then explored the performance of those PTMs
for modelling turbulent transport, using the Hasegawa-Wakatani system
as a minimal model for 2D instability-driven turbulence for tokamak
plasmas.

In fixed-gradient simulations, we observe that keeping at least 4
poloidal modes, distributed around the most unstable mode, was sufficient
to observe a sharp transition from turbulence to ZFs as the linear
parameter $C/\kappa$ is scanned, as well as the known scalings of
the particle flux in these two regimes, although the ZF fraction is
overestimated in the turbulent regime. We find that, using at least
10 modes, it was possible to obtain a reliable description of both
turbulent and ZF dominated states, albeit a shifted transition point,
with respect to DNS.

Turning to the flux-driven HW system, again 4 modes were found to
be the bare minimum to correctly describe both turbulent and ZF states,
with the same time evolution for the mean density gradient. Furthermore,
we found that we get almost the same PDF as in DNS for the particle
flux, with 10 modes, which suggests that such a model can even describe
some of the statistics of the stationary state.

Analysing fixed-gradient simulations, in the turbulent regime, we
observed inverse energy and direct enstrophy cascades consistent with
the dual cascade picture of 2D Navier-Stokes turbulence. In the ZF
dominated regime, large poloidal scales $k_{y}<k_{y0}/2$ remove energy
from ZFs, while smaller scales $k_{y}\gtrsim k_{y0}$, \emph{including
the linearly most unstable mode}, feed the ZFs. In that case, the
dual cascade picture still holds, but with an anisotropic inverse
energy transfer from the injection at the most unstable mode, through\emph{
}ZFs, and finally to large poloidal scales. The direct enstrophy cascade,
on the other end, stays isotropic. This provides a necessary selection criterion
for building PTMs, which should include poloidal scales both larger
and smaller than the most unstable mode $k_{y0}$ in order to correctly describe
these different transfers and interactions, which is achieved here with $n_y\geq4$.

What is not captured yet by the models is the dynamics close to the
transition point $C/\kappa\approx0.1$, the hysteresis loop around
this threshold, and consequently profile stiffness in the flux-driven
framework. Understanding what are the missing ingredients, related
to the comprehension of the transition mechanism, is the goal of our
future works. Designing statistical closures \citep{krommes:2002},
or closures in the spirit of LES \citep{morel:2012}, will be also
tackled in the future, along with considering other reductions, including
hybrid lattices \citep{gurcan:2026} or wavelets \citep{farge:1992}.

Finally, we underline that reduced fluid models, like the HW system,
enable to easily explore such reductions, and to perform the kind
of analysis as the one presented here. Detailed studies of such reductions
should also elucidate the effect of having too low a resolution, or
a non-standard domain aspect ratio, on the system, and particularly
on the self-organisation of turbulence. But, eventually, PTMs, or
a similar methodology, could also be potentially extended to more
complex systems, i.e. adding interchange, turning to fluid ITG, and
finally to gyrofluid or gyrokinetic systems that incorporate temperature
profile evolution, multi-species effects, alpha-particle physics,
helium-ash transport, electromagnetic fluctuations, and realistic
toroidal geometry.
\begin{acknowledgments}
This work was granted access to the Jean Zay super-computer of IDRIS
under the allocation AD010514291R2 by GENCI, and has been carried
out within the framework of the EUROfusion Consortium, funded by the
European Union \emph{via} the Euratom Research and Training Programme
(Grant Agreement No 101052200 --- EUROfusion) and within the framework
of the French Research Federation for Fusion Studies. The authors
would like to thank the Isaac Newton Institute for Mathematical Sciences,
Cambridge, for support and hospitality during the programme ``Anti-diffusive
dynamics: from sub-cellular to astrophysical scales''. The authors
acknowledge stimulating discussions with participants at the 2025
Festival de Théorie in Aix-en-Provence. PLG thanks L. Manfredini for
his help and intuitions on energy transfers between clusters of Fourier
modes.
\end{acknowledgments}

\appendix

\section{Eigenvalues of the Hasegawa-Wakatani system}

\label{sec:app-eigenvalues}

The linearised Fourier transform of the HW equations (\ref{eq:hw})
for non-zonal modes $(k_{y}\neq0)$ can be rewritten \citep{gurcan:2022}
\begin{subequations}
\begin{align}
\partial_{t}\phi_{k}+(A_{k}-B_{k})\phi_{k} & =\frac{C}{k^{2}}n_{k}\,,\label{eq:HWappphik}\\
\partial_{t}n_{k}+(A_{k}+B_{k})n_{k} & =(C-i\kappa k_{y})\phi_{k}\text{ ,}\label{eq:HWappnk}
\end{align}
\end{subequations}
 where 
\begin{subequations}
\begin{align}
A_{k}= & \frac{1}{2}\left[(Dk^{2}+C)+\left(\frac{C}{k^{2}}+\nu k^{2}\right)\right],\label{eq:HWappAk}\\
B_{k}= & \frac{1}{2}\left[(Dk^{2}+C)-\left(\frac{C}{k^{2}}+\nu k^{2}\right)\right].\label{eq:HWappBk}
\end{align}
\end{subequations}

The two eigenvalues $\omega^{\pm}_{k}=\omega^{\pm}_{k,r}+i\gamma^{\pm}_{k}$
are 
\begin{equation}
\omega^{\pm}_{k}=\pm W_{k}-iA_{k},\label{eq:omkapp}
\end{equation}
with 
\begin{equation}
W_{k}=\frac{k_{y}}{|k_{y}|}\sqrt{\frac{H_{k}-G_{k}}{2}}+i\sqrt{\frac{H_{k}+G_{k}}{2}},\label{eq:Omkpmapp}
\end{equation}
and 
\begin{equation}
H_{k}=\sqrt{G^{2}_{k}+C^{2}\kappa{^{2}}k^{2}_{y}/k^{4}},\quad G_{k}=\left(B^{2}_{k}+\frac{C^{2}}{k^{2}}\right).\label{eq:GHkapp}
\end{equation}

\begin{figure}[tbph]
\centering{}\textcolor{violet}{\includegraphics[width=1\columnwidth]{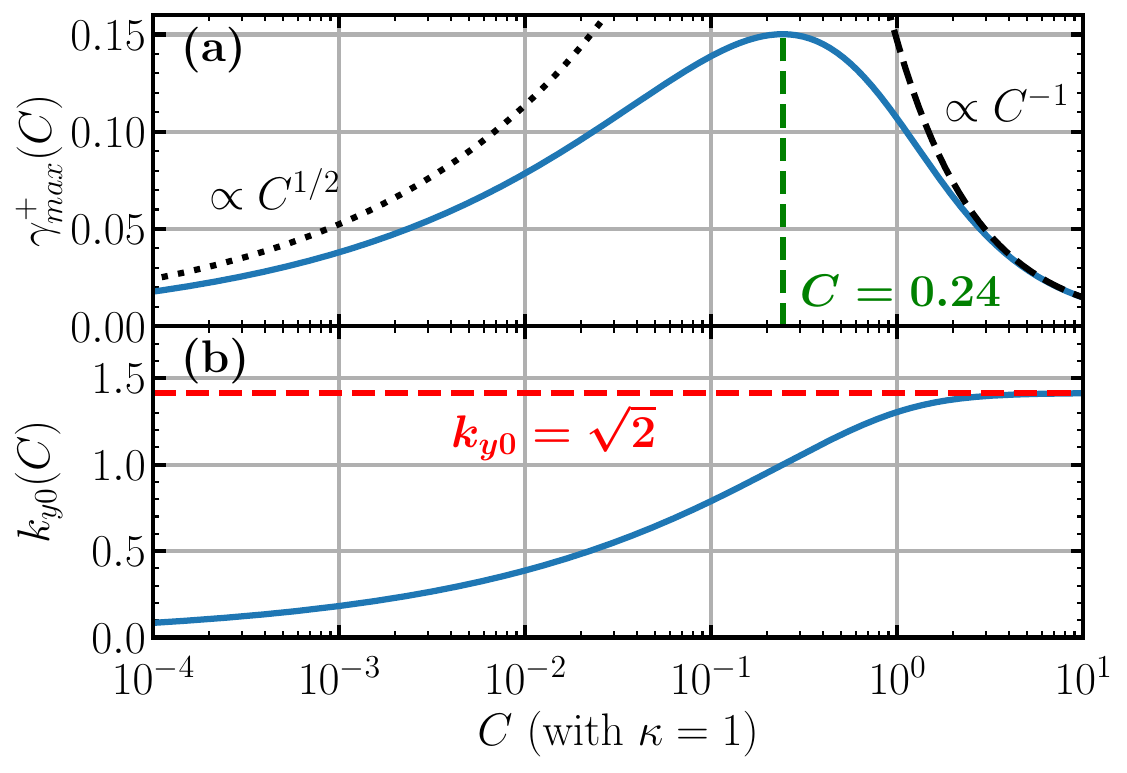}}\caption{Most unstable growth rate $\gamma^{+}_{max}$ (a) and most unstable
wave-number $k_{y0}$ (b) as functions of the adiabaticity parameter
$C$ (in the inviscid case and with $\kappa=1$). The analytical solutions
in both hydrodynamic ($C\ll1$, $\gamma^{+}_{max}\propto C^{1/2}$)
and adiabatic ($C\gg1$, $\gamma^{+}_{max}\propto C^{-1}$) are shown
in dotted and dashed line, respectively.}
\label{fig:gamky0C}
\end{figure}

The most unstable growth rate $\gamma^{+}_{max}$ and the corresponding
most unstable wave-number $k_{y0}$ are shown as functions of the
adiabaticity parameter $C$ in Figure \ref{fig:gamky0C} (a) and (b).

\section{Other configurations}

\label{sec:Other-reductions}

In this Section, we discuss the other configurations for PTMs that
we tried in the fixed gradient case and which did not provide satisfactory
or interesting results.

\subsection{Intermediary reductions}

\begin{figure}[tbph]
\centering{}\includegraphics[width=.9\columnwidth]{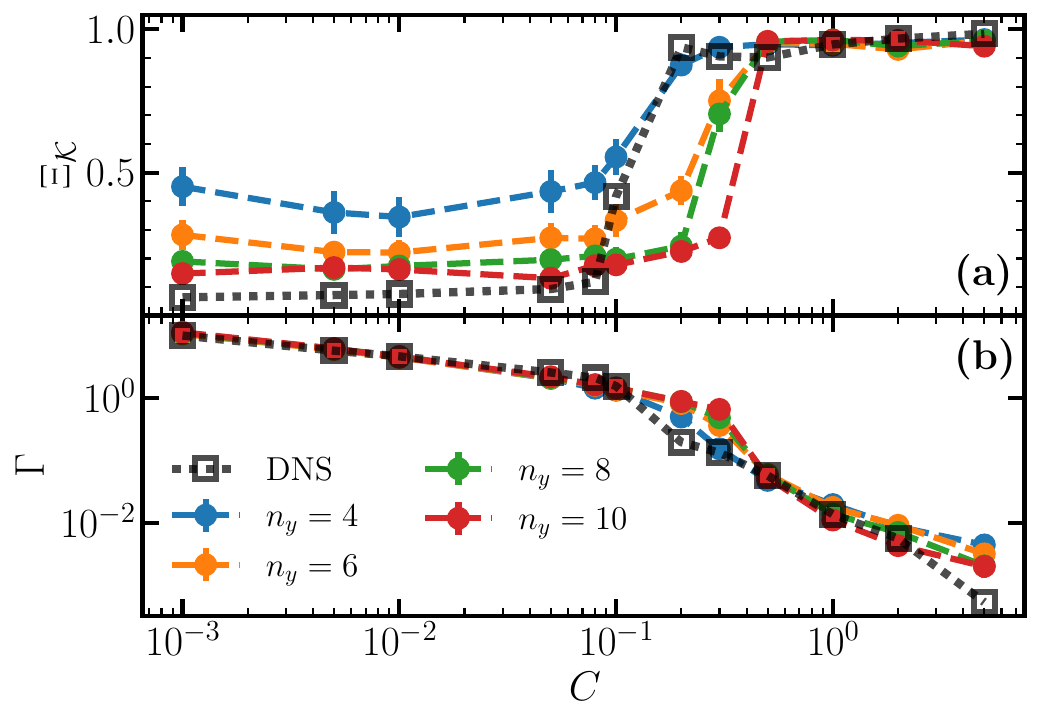}\caption{ZF level $\Xi_{\mathcal{K}}=\mathcal{K}_{ZF}/\mathcal{K}$ (a) and
mean radial particle flux $\Gamma$ (b) versus adiabaticity parameter
$C$ for DNS (black), and PTMs from $n_{y}=4$ to $n_{y}=10$. Both
quantities are time averaged over the second half of the simulations.}
\label{fig:Cscan-intermediary}
\end{figure}

\begin{table*}[tbph]
\centering{}%
\begin{tabular*}{1\textwidth}{@{\extracolsep{\fill}}@{\extracolsep{\fill}}lcccc}
\hline 
Poloidal resolution $n_{y}$ and label & 10, $1k$ & 10, $3k$ & 20, $4k$, $\nu_{y}=\nu_{x}$ & 41, $4k$, $\nu_{y}=\nu_{x}$\tabularnewline
\hline 
Selected wave-numbers $k^{j}_{y}$ for $j\in[1,n_{y}]$ & $jk_{y0}/10$ & $3jk_{y0}/10$ & $jk_{y0}/5$ & $jk_{y0}/10$\tabularnewline
Dissipation along $k_{y}$ & 0 & 0 & $0.02\cdot\gamma^{+}_{max}/k_{y0}$ & $0.02\cdot\gamma^{+}_{max}/k_{y0}$\tabularnewline
pLES & yes & no & yes & yes\tabularnewline
Corresponding padded resolution & $1024\times34$ & $1024\times34$ & $1024\times64$ & $1024\times128$\tabularnewline
\hline 
\end{tabular*}\caption{Selected poloidal wave-numbers and corresponding padded resolution
for each PTM with poloidal resolution $n_{y}$ (and their label on
Figure \ref{fig:Cscan-other}). The flag \textquotedblleft pLES\textquotedblright{}
indicates if the reduced model corresponds to a poloidal LES.}
\label{tab:PTRMotherselect}
\end{table*}

We consider the three following reductions: $n_{y}=4$, $n_{y}=6$
and $n_{y}=8$, where we select the modes $2jk_{y0}/n_{y}$ for $j\in[1,n_{y}]$,
such that we keep modes from $2k_{y0}/n_{y}$ to $2k_{y0}$, as for
$n_{y}=10$ and $n_{y}=20$ PTMs studied in the article. The simulations
are performed as in Section \ref{sec:fixedgrad}, and the results
are shown in Figure \ref{fig:Cscan-intermediary}, where we also plot
the results for $n_{y}=10$ (red) and the DNS (black). First, we notice
an upward shift accompanied by a sharpening of the transition as we
increase the resolution. Hence, while $n_{y}=4$ (blue) predict the
good transition point, the corresponding transition is too smooth
and the level of ZFs too high in the turbulent regime (which makes
sense because $1/4$ of the reduced grid is composed of zonal modes.
On the contrary, $n_{y}=8$ (green) and $n_{y}=10$ are quite similar,
the latter being slightly sharper, but the transition point is overestimated
at $C\approx0.3$. Note that the ZF velocity profiles (not shown)
are all very similar to DNS, and not as noisy as for $n_{y}=1$ or
$n_{y}=2$ as previously discussed in Section \ref{sec:fixedgrad}.

\subsection{Other reductions and modes selection}

\begin{figure}[tbph]
\centering{}\includegraphics[width=.9\columnwidth]{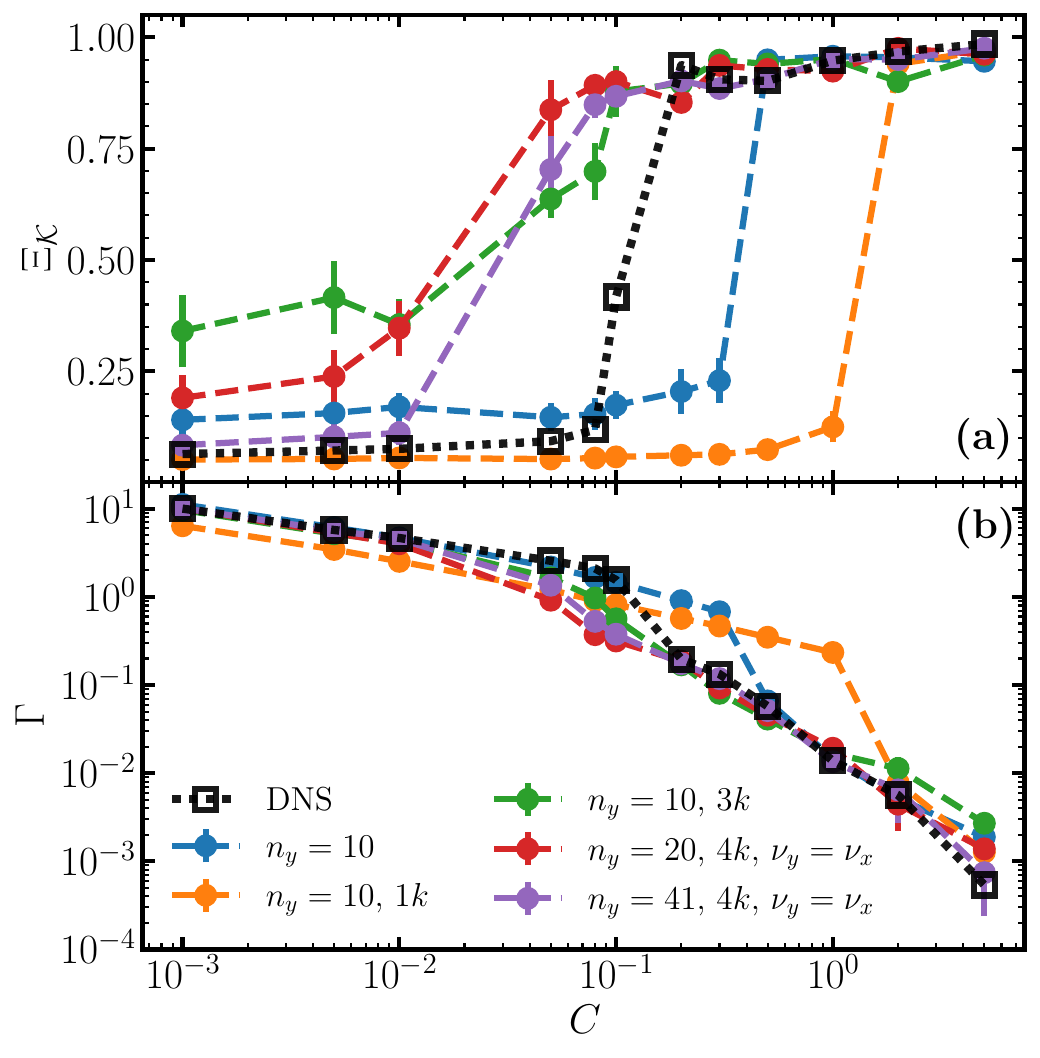}\caption{ZF level $\Xi_{\mathcal{K}}$ (a) and mean flux $\Gamma$ (b) versus
$C$ for DNS (black), and PTMs given in Table \ref{tab:PTRMotherselect}.}
\label{fig:Cscan-other}
\end{figure}

The other configurations are shown in Table \ref{tab:PTRMotherselect},
and the results in Figure \ref{fig:Cscan-other}. Particularly, using
$n_{y}=10$ but with modes $k^{i}_{y}=ik_{y0}/10\in[k_{y0}/10,k_{y0}]$,
for $i\in[1,10]$, we find that the transition point is shifted to
$C=1$. Conversely, reaching smaller poloidal scales with $n_{y}=10$
($k^{i}_{y}\in[3k_{y0}/10,3k_{y0}]$), or with $n_{y}=20$ ($k^{i}_{y}\in[k_{y0}/5,4k_{y0}]$)
leads to an overestimation of the ZF level in the turbulent regime,
and a shift of the transition point towards lower $C$ values.

\section{Impact of dissipation on the transition}

\label{sec:Diss}

\begin{figure}[tbph]
\centering{}\includegraphics[width=.9\columnwidth]{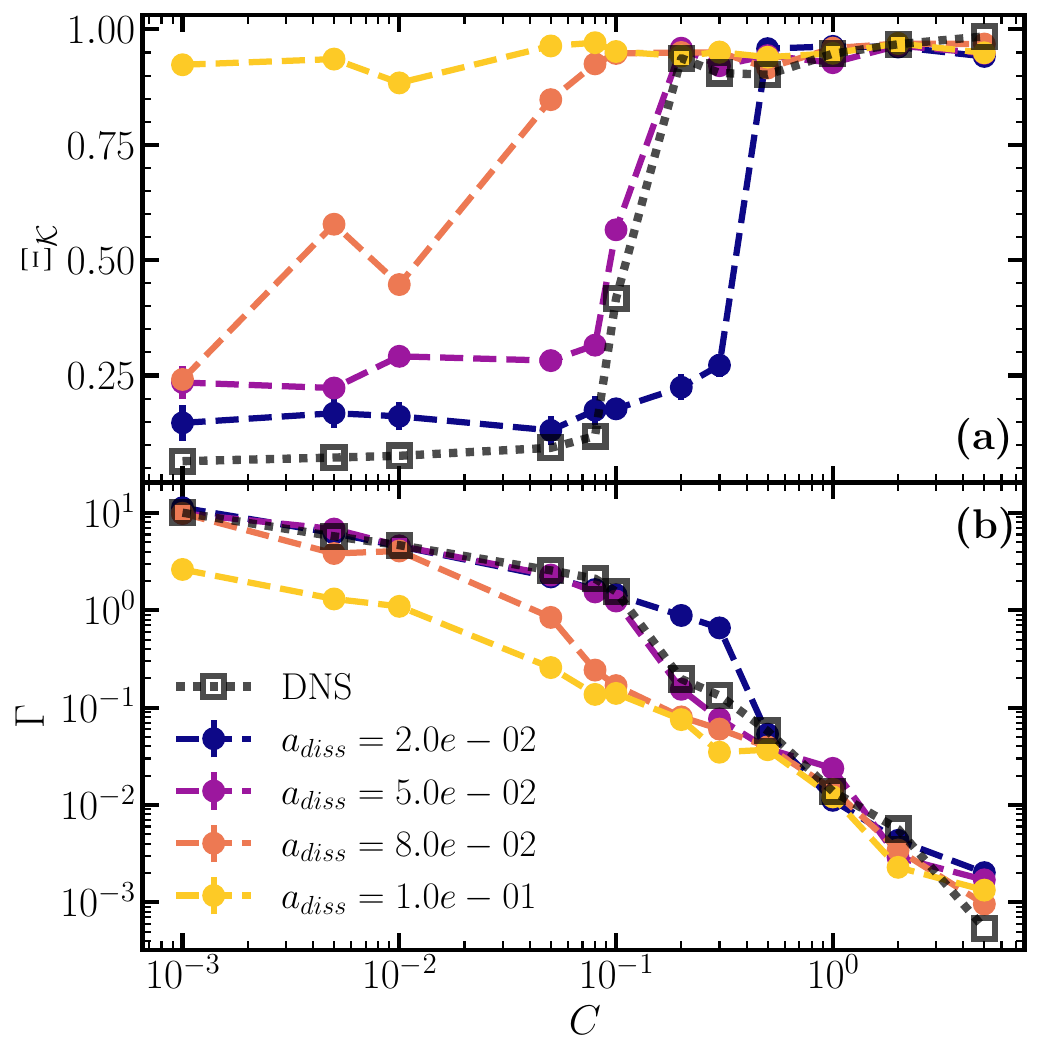}\caption{ZF level $\Xi_{\mathcal{K}}$ (a) and mean flux $\Gamma$ (b) versus
$C$ for DNS (black), and PTM $n_{y}=10$ with increasing factor $a_{diss}$.}
\label{fig:Cscan-diss}
\end{figure}

To study the dependency on dissipation of the transition for the PTM
$n_{y}=10$, we set $\nu(C)=D(C)=a_{diss}\gamma^{+}_{max}(C)/k^{2}_{y0}(C)$
and perform the same $C$ scan for different values of $a_{diss}$
($a_{diss}=2e-2$ in Section \ref{sec:fixedgrad}, see Table \ref{tab:GeneralDNS}).
The results are shown in Figure \ref{fig:Cscan-diss} for $a_{diss}=2e-2$
(blue), $a_{diss}=5e-2$ (purple), $a_{diss}=8e-2$ (orange) and $a_{diss}=1e-1$
(yellow), and compared to DNS (black dotted line and squares).

While slightly increasing $a_{diss}$ to $5e-2$ allows to recover
the correct transition point $C=0.1$ and the appropriate particle
flux (although the ZF level also increases in the turbulence regime),
further increasing $a_{diss}$ progressively erases the transition,
until the system is always ZF dominated, for $a_{diss}=1e-1$, regardless
of $C$. This highlights both the sensitivity of the reduced models
on the dissipation, and the fact that using higher dissipation as
a closure is not appropriate.

\section{Penalisation parameters (flux-driven)}

\label{sec:Penalisation-parameters}

The parameters for the penalisation method used in the code P-FLARE
are given in Table \ref{tab:paramspen} for the flux-driven simulations
in Section \ref{sec:fluxdriven}, taken as in Table 3 from Ref. \citealp{guillon:2025c},
were the methodology can be found.

\begin{table*}[tbph]
\begin{centering}
\begin{tabular*}{1\textwidth}{@{\extracolsep{\fill}}@{\extracolsep{\fill}}cccccccc}
\toprule 
\multicolumn{3}{c}{Buffer zone} & \multicolumn{3}{c}{Smoothing function} & Penalisation coefficient & Artificial source size\tabularnewline
\midrule 
$x_{b1}$ & $x_{b2}$ & $\delta x_{b}$ & $x_{m1}$ & $x_{m2}$ & $\delta x_{m}$ & $\mu$ & $\sigma_{S_{b}}$\tabularnewline
$90\Delta x$ & $L_{x}-90\Delta x$ & $60\Delta x$ & $45\Delta x$ & $L_{x}-45\Delta x$ & $40\Delta x$ & 100 & $5L_{x}$\tabularnewline
\bottomrule
\end{tabular*}
\par\end{centering}
\caption{Penalisation parameters for both DNS and PTM simulations in Section
\ref{subsec:Flux-driven-farmargi}. The radial grid increment is $\Delta x=L_{x}/N_{x}$,
where $N_{x}=2\lfloor N_{px}/3\rfloor$ is the unpadded resolution,
$N_{px}=1024$, and $\Delta x$ changes with $C$, by definition of
$L_{x}$ in Table \ref{tab:GeneralDNS}.}
\label{tab:paramspen}
\end{table*}

\section{Exponentially modified Gaussian}

\label{sec:EMG}

Decomposing the radial particle flux $\Gamma_{n}$ into two parts
$\Gamma_{n}=\Gamma_{\mathcal{N}}+\Gamma_{\mathcal{E}}$, where $\Gamma_{\mathcal{N}}$
follows a normal law $\mathcal{N}(\mu,\sigma)$ with mean $\mu$ and
standard deviation $\sigma$, and $\Gamma_{\mathcal{E}}$ follows
an exponential law $\mathcal{E}(\lambda)$ of mean $1/\lambda$, it
can be shown that $\Gamma_{n}$ follows an EMG of parameters $\left(\mu,\sigma,\lambda\right)$,
of distribution function \citep{dyson:1998a,:m} 
\begin{align}
p(\Gamma)= & \frac{\lambda}{2}\exp\left[\frac{\lambda}{2}\left(2\mu+\lambda\sigma^{2}-2\Gamma\right)\right]\text{erfc}\left[\frac{\mu+\lambda\sigma^{2}-\Gamma}{\sqrt{2}\sigma}\right],\label{eq:pdfEMG}
\end{align}
where $\text{erfc}=1-\text{erf}$, with $\text{erf}$ the error function.
Eq. \ref{eq:pdfEMG} is obtained by taking the convolution product
between the normal and the exponential distributions. The EMG law
thus describes a process which has a symmetric Gaussian part and a
positively skewed exponential-law part. Its mean $\mu_{EMG}$, standard
deviation $\sigma_{EMG}$ and skewness $\text{sk}_{EMG}$ are given
by $\mu_{EMG}=\mu+1/\lambda$, $\sigma^{2}_{EMG}=\sigma^{2}+1/\lambda^{2}$
and $\text{sk}_{EMG}=2\left(1+\sigma^{2}\lambda^{2}\right)^{-3/2}$,
so that $\mu,\sigma,\lambda$ can be obtained from the particle flux
PDFs using the flux-driven DNS from Section \ref{sec:fluxdriven}.

\begin{figure}[tbph]
\centering{}\includegraphics[width=1\columnwidth]{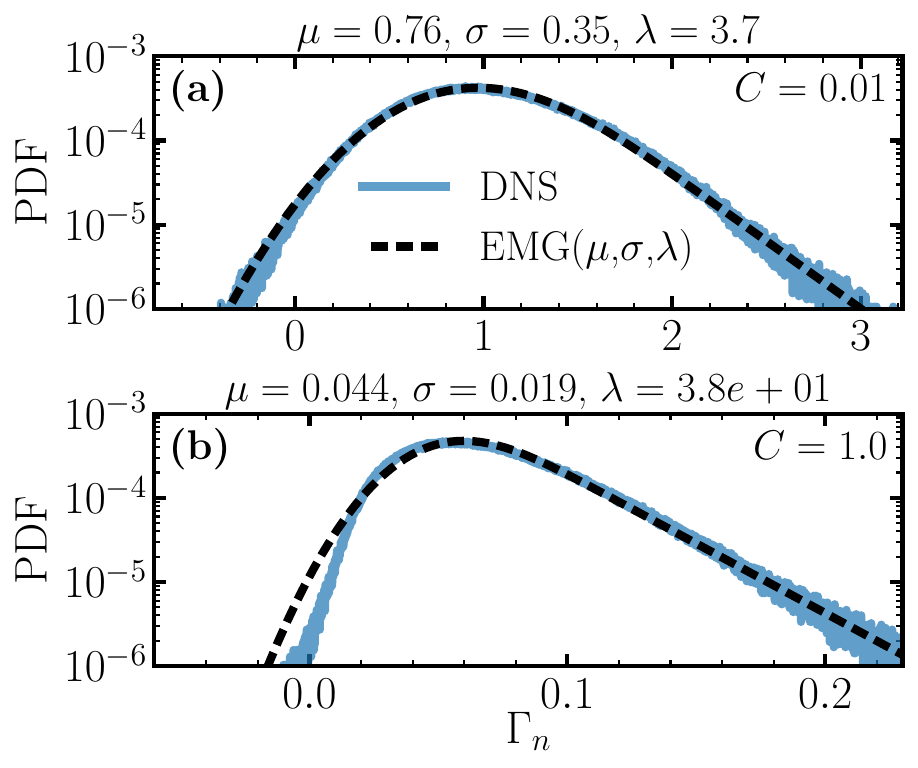}\caption{PDF of $\Gamma_{n}$, normalised by its sum, for the flux-driven DNS
$C=0.01$ (a) and $C=1.0$ (b), computed as in Section \ref{sec:fluxdriven}.
The EMG PDF (normalised by its sum) is the black dashed line.}
\label{fig:DNSfluxpdf}
\end{figure}

The results are shown in Figure \ref{fig:DNSfluxpdf}, where the EMG
PDFs, determined from the simulations, are shown in black dashed lines.
For the turbulent case $C=0.01$ (a), there is a very good agreement
with DNS, suggesting that the flux might be decomposed between a Gaussian
and an exponential part. The former may model bulk turbulence, while
the latter could represent the emission of large scale coherent structures
in the outward direction. In the ZF dominated case $C=1$ (b), the
agreement is less correct and the PDF from DNS is narrower and more
skewed. This could be related to barriers of transport represented
by ZFs, which only let strongly coherent structures, corresponding
to rarer events, pass.

\section{Quantitative comparison between DNS and PTMs in flux-driven}\label{sec:quant}

In Table \ref{tab:relativeFD}, we show the relative error of the
ZF energy fraction, the background density gradient and of the mean
particle flux, averaged over time, between each PTM and the DNS, in
both turbulent (top) and zonal flow (bottom) cases. The relative errors
are compared to the intrinsic relative error of the DNS, which is
the ratio between the standard deviation and the mean value of each
measured quantity.

\begin{table*}[htbp]
\subfloat[Turbulent case $C=0.01$]{%
\begin{tabular*}{1\textwidth}{@{\extracolsep{\fill}}lccccc}
\toprule
Relative errors & DNS (self) & $n_{y}=10$ & $n_{y}=4$ & $n_{y}=2$ & $n_{y}=1$\tabularnewline
\hline 
Zonal fraction & 33\% & 34\% & 107\% & 51\% & 59\%\tabularnewline
Background gradient & 1\% & 3\% & 4\% & 11\% & 11\%\tabularnewline
Mean flux & 11\% & 14\% & 17\% & 26\% & 39\%\tabularnewline
\bottomrule
\end{tabular*}

}

\subfloat[Zonal flow dominated case $C=1$]{%
\begin{tabular*}{1\textwidth}{@{\extracolsep{\fill}}lccccc}
\toprule
Relative errors & DNS (self) & $n_{y}=10$ & $n_{y}=4$ & $n_{y}=2$ & $n_{y}=1$\tabularnewline
\hline 
Zonal fraction & 0.6\% & 2\% & 3\% & 26\% & 28\%\tabularnewline
Background gradient & 3\% & 0.3\% & 1\% & 16\% & 15\%\tabularnewline
Mean flux & 14\% & 17\% & 21\% & 151\% & 152\%\tabularnewline
\bottomrule
\end{tabular*}

}

\caption{Time-averaged relative errors of zonal flow level, mean density gradient
and mean particle flux, between each PTM and DNS, for both turbulent
(top table) and zonal flow dominated (bottom table) flux-driven simulations.
The relative errors are compared with the intrinsic relative error
of the DNS, which is the ratio between the standard deviation and
the mean value of each measured quantity.}\label{tab:relativeFD}
\end{table*}

The overall observation is that, in both regimes, only $n_{y}=4$,
and better $n_{y}=10$, have relative errors for all quantities that
are of the same order, or smaller than the DNS intrinsic relative
error. The only exception is the zonal fraction for $n_{y}=4$ in
the turbulent regime, which is almost 2 times the fraction measured
in DNS, due to the large number of zonal modes compared to the total
number of modes, and which could be corrected by applying some weights.

In Table \ref{tab:momentsFD}, we then compare the mean, standard
deviation, skewness and kurtosis of the particle flux PDF computed
for the DNS and each PTM. Although the mean values are of the similar
order of magnitude for each reduced model, the even order moments
for $n_{y}=1$ and $n_{y}=2$ are much larger compared to DNS, than
those of $n_{y}=4$ and $n_{y}=10$, which is related to the lowest
resolution models having too flat PDFs corresponding to a much more
intermittent and abrupt transport.

\begin{table*}[htbp]
\subfloat[Turbulent case $C=0.01$]{%
\begin{tabular*}{1\textwidth}{@{\extracolsep{\fill}}lccccc}
\toprule
PDF Measurements & DNS & $n_{y}=10$ & $n_{y}=4$ & $n_{y}=2$ & $n_{y}=1$\tabularnewline
\hline 
Mean & 1.04 & 1.00 & 0.98 & 0.82 & 0.91\tabularnewline
Standard deviation & 0.45 & 0.67 & 1.03 & 1.21 & 2.23\tabularnewline
Skewness & 0.44 & 0.34 & 0.50 & 0.65 & 0.86\tabularnewline
Kurtosis & 3.42 & 3.42 & 4.06 & 5.09 & 7.55\tabularnewline
Wasserstein-1 distance with DNS & - & 0.18 & 0.45 & 0.61 & 1.26\tabularnewline
\bottomrule
\end{tabular*}

}

\subfloat[Zonal flow dominated case $C=1$]{%
\begin{tabular*}{1\textwidth}{@{\extracolsep{\fill}}lccccc}
\toprule
PDF Measurements & DNS & $n_{y}=10$ & $n_{y}=4$ & $n_{y}=2$ & $n_{y}=1$\tabularnewline
\hline 
Mean & 0.070 & 0.066 & 0.064 & 0.162 & 0.165\tabularnewline
Standard deviation & 0.033 & 0.035 & 0.054 & 0.164 & 0.209\tabularnewline
Skewness & 1.04 & 1.04 & 1.71 & 1.73 & 2.15\tabularnewline
Kurtosis & 4.88 & 4.64 & 7.61 & 8.72 & 12.90\tabularnewline
Wasserstein-1 distance with DNS & - & 0.004 & 0.018 & 0.11 & 0.124\tabularnewline
\bottomrule
\end{tabular*}

}

\caption{Comparison of the mean, standard deviation, skewness and kurtosis
for the radial flux PDF between DNS and the reduced model, for both
turbulent (top table) and zonal flow dominated (bottom table) flux-driven
simulations. The Wasserstein-1 distance between the PDFs in each PTM
and that of DNS is shown at the bottom line.}\label{tab:momentsFD}
\end{table*}
 We also compute the Wasserstein-1 distance between the PDF of each
reduced model and that of DNS. Overall, the distance for $n_{y}=1$
and $n_{y}=2$ is of the same order of magnitude of the mean flux
in DNS, while it is smaller for $n_{y}=4$ and even more so for $n_{y}=10$.
This highlights that $n_{y}=10$ exhibits PDFs that are close to that
of DNS, compared to the other models.

\section{Clustering method}

\label{sec:CLUST}

The time evolution of the kinetic energy $e_{k}=\frac{1}{2}k^{2}|\phi_{k}|^{2}$
at the mode $k$ is given by 
\begin{equation}
\partial_{t}e_{k}=\text{Re}\left[\phi^{*}_{k}\mathcal{F}_{k}\left\{ [\phi,\nabla^{2}\phi]\right\} \right]=\sum_{k+p+q=0}T_{kpq},\label{eq:dtek}
\end{equation}
where $T_{kpq}$ is the energy transfer to $k$ in the triad that
satisfy $k+p+q=0$, and $\mathcal{F}_{k}\left\{ [\phi,\nabla^{2}\phi]\right\} $
is the Fourier transform of the non-linear term in Eq. \ref{eq:hw1}.
Let us denote the clusters of Fourier modes by $\mathcal{V}_{i}$.
To compute the transfer from cluster $\mathcal{V}_{i}$ to cluster
$\mathcal{V}_{j}$ mediated by $\mathcal{V}_{\ell}$, noted $\mathcal{T}^{\ell}_{i\to j}$,
we have two possibilities. If $\ell=i$ or $\ell=j$, then we just
sum over the triads $\left(k_{i},k_{j},k_{\ell}\right)\in\mathcal{V}_{i}\times\mathcal{V}_{j}\times\mathcal{V}_{\ell}$:
\begin{align}
\mathcal{T}^{\ell\in\{i,j\}}_{i\to j} & =\sum_{k_{i}\in\mathcal{V}_{i}}\sum_{k_{j}\in\mathcal{V}_{j}}\sum_{k_{\ell}\in\mathcal{V}_{\ell}}T_{k_{i}k_{j}k_{\ell}}\delta_{ij\ell},\label{eq:Titojmedl}
\end{align}
where $\delta_{ij\ell}=\delta\left(k_{i}+k_{j}+k_{\ell}\right)$.
If now $\ell$ corresponds to a third cluster, different from $i$
and $j$, we compute for each triad $\left(k_{i},k_{j},k_{\ell}\right)$
the transfer from $k_{i}$ to $k_{j}$ \emph{mediated} by $k_{\ell}$
\citep{gurcan:2023} 
\begin{equation}
t^{k_{\ell}}_{k_{i}\to k_{j}}=\frac{1}{3}\left(T_{k_{i}k_{j}k_{\ell}}-T_{k_{j}k_{\ell}k_{i}}\right),\label{eq:tijlmed}
\end{equation}
and we sum over all modes to obtain 
\begin{align}
\mathcal{T}^{\ell\notin\{i,j\}}_{i\to j} & =\frac{1}{3}\sum_{k_{i},k_{j}k_{\ell}}\left(T_{k_{i}k_{j}k_{\ell}}-T_{k_{j}k_{\ell}k_{i}}\right)\delta_{ij\ell}.\label{eq:Titojdiffl}
\end{align}
The total transfer from $i$ to $j$ is thus 
\begin{equation}
\mathcal{T}_{i\to j}=\mathcal{T}^{\ell\in\{i,j\}}_{i\to j}+\sum_{\ell\notin\{i,j\}}\mathcal{T}^{\ell\notin\{i,j\}}_{i\to j},\label{eq:Titoj}
\end{equation}
for an arbitrary number of clusters $\ell\notin\{i,j\}$. For enstrophy,
we use $T_{kpq}\to k^{2}T_{kpq}$. In practice, we compute the transfers
using $\mathcal{F}_{k}\left\{ -[\phi,\nabla^{2}\phi]\right\} $ from
simulations, by filtering $\phi$ and $\nabla^{2}\phi$ on the different
clusters.

\begin{thebibliography}{10}

\bibitem{hasegawa:1983}
A.~Hasegawa and M.~Wakatani.
\newblock Plasma {{Edge Turbulence}}.
\newblock {\em Phys. Rev. Lett.}, 50(9):682--686, February 1983.
\newblock \href {https://doi.org/10.1103/PhysRevLett.50.682}
  {\path{doi:10.1103/PhysRevLett.50.682}}.

\bibitem{numata:2007}
R.~Numata, R.~Ball, and R.~L. Dewar.
\newblock Bifurcation in electrostatic resistive drift wave turbulence.
\newblock {\em Physics of Plasmas}, 14(10):102312, October 2007.
\newblock \href {https://doi.org/10.1063/1.2796106}
  {\path{doi:10.1063/1.2796106}}.

\bibitem{guillon:2025}
P.~L. Guillon and {\"O}.~D. G{\"u}rcan.
\newblock Phase transition from turbulence to zonal flows in the
  {{Hasegawa}}--{{Wakatani}} system.
\newblock {\em Physics of Plasmas}, 32(1):012306, January 2025.
\newblock \href {https://doi.org/10.1063/5.0242282}
  {\path{doi:10.1063/5.0242282}}.

\bibitem{guillon:2025d}
P.~L. Guillon, G.~{Dif-Pradalier}, Y.~Sarazin, D.~W. Hughes, and {\"O}.~D.
  G{\"u}rcan.
\newblock Self-organisation through layering of {$\beta$}-plane like turbulence
  in plasmas and geophysical fluids.
\newblock {\em accepted to Philos. Trans. R. Soc. A}, November 2025.
\newblock \href {https://doi.org/10.48550/arXiv.2511.10438}
  {\path{doi:10.48550/arXiv.2511.10438}}.

\bibitem{garbet:1998}
X.~Garbet and R.~E. Waltz.
\newblock Heat flux driven ion turbulence.
\newblock {\em Physics of Plasmas}, 5(8):2836--2845, August 1998.
\newblock \href {https://doi.org/10.1063/1.873003}
  {\path{doi:10.1063/1.873003}}.

\bibitem{gillot:2023}
C.~Gillot, G.~{Dif-Pradalier}, Y.~Sarazin, C.~Bourdelle, A.~B. Navarro,
  Y.~Camenen, J.~Citrin, A.~D. Siena, X.~Garbet, P.~Ghendrih, V.~Grandgirard,
  P.~Manas, and F.~Widmer.
\newblock The problem of capturing marginality in model reductions of
  turbulence.
\newblock {\em Plasma Phys. Control. Fusion}, 65(5):055012, March 2023.
\newblock \href {https://doi.org/10.1088/1361-6587/acc276}
  {\path{doi:10.1088/1361-6587/acc276}}.

\bibitem{biglari:1990}
H.~Biglari, P.~H. Diamond, and P.~W. Terry.
\newblock Influence of sheared poloidal rotation on edge turbulence.
\newblock {\em Physics of Fluids B: Plasma Physics}, 2(1):1--4, January 1990.
\newblock \href {https://doi.org/10.1063/1.859529}
  {\path{doi:10.1063/1.859529}}.

\bibitem{hinton:1991}
F.~L. Hinton.
\newblock Thermal confinement bifurcation and the {{L}}- to {{H}}-mode
  transition in tokamaks.
\newblock {\em Phys. Fluids B}, 3(3):696--704, March 1991.
\newblock \href {https://doi.org/10.1063/1.859866}
  {\path{doi:10.1063/1.859866}}.

\bibitem{hinton:1993}
F.~L. Hinton and G.~M. Staebler.
\newblock Particle and energy confinement bifurcation in tokamaks.
\newblock {\em Physics of Fluids B: Plasma Physics}, 5(4):1281--1288, April
  1993.
\newblock \href {https://doi.org/10.1063/1.860919}
  {\path{doi:10.1063/1.860919}}.

\bibitem{diamond:1995}
P.~H. Diamond and T.~S. Hahm.
\newblock On the dynamics of turbulent transport near marginal stability.
\newblock {\em Physics of Plasmas}, 2(10):3640--3649, October 1995.
\newblock \href {https://doi.org/10.1063/1.871063}
  {\path{doi:10.1063/1.871063}}.

\bibitem{carreras:1996}
B.~A. Carreras, D.~Newman, V.~E. Lynch, and P.~H. Diamond.
\newblock A model realization of self-organized criticality for plasma
  confinement.
\newblock {\em Physics of Plasmas}, 3(8):2903--2911, August 1996.
\newblock \href {https://doi.org/10.1063/1.871650}
  {\path{doi:10.1063/1.871650}}.

\bibitem{diamond:2001}
P.~H. Diamond, S.~Champeaux, M.~Malkov, A.~Das, I.~Gruzinov, M.~N. Rosenbluth,
  C.~Holland, B.~Wecht, A.~I. Smolyakov, F.~L. Hinton, Z.~Lin, and T.~S. Hahm.
\newblock Secondary instability in drift wave turbulence as a mechanism for
  zonal flow and avalanche formation.
\newblock {\em Nucl. Fusion}, 41(8):1067, August 2001.
\newblock \href {https://doi.org/10.1088/0029-5515/41/8/310}
  {\path{doi:10.1088/0029-5515/41/8/310}}.

\bibitem{diamond:2005}
P.~H. Diamond, S.-I. Itoh, K.~Itoh, and T.~S. Hahm.
\newblock Zonal flows in plasma---a review.
\newblock {\em Plasma Phys. Control. Fusion}, 47(5):R35, April 2005.
\newblock \href {https://doi.org/10.1088/0741-3335/47/5/R01}
  {\path{doi:10.1088/0741-3335/47/5/R01}}.

\bibitem{dif-pradalier:2015}
G.~{Dif-Pradalier}, G.~Hornung, {\relax Ph}.~Ghendrih, Y.~Sarazin, F.~Clairet,
  L.~Vermare, P.~H. Diamond, J.~Abiteboul, T.~{Cartier-Michaud}, C.~Ehrlacher,
  D.~Est{\`e}ve, X.~Garbet, V.~Grandgirard, {\"O}.~D. G{\"u}rcan, P.~Hennequin,
  Y.~Kosuga, G.~Latu, P.~Maget, P.~Morel, C.~Norscini, R.~Sabot, and
  A.~Storelli.
\newblock Finding the {{Elusive ExB Staircase}} in {{Magnetized Plasmas}}.
\newblock {\em Phys. Rev. Lett.}, 114(8):085004, February 2015.
\newblock \href {https://doi.org/10.1103/PhysRevLett.114.085004}
  {\path{doi:10.1103/PhysRevLett.114.085004}}.

\bibitem{gurcan:2015}
{\"O}.~D. G{\"u}rcan and P.~H. Diamond.
\newblock Zonal flows and pattern formation.
\newblock {\em J. Phys. A: Math. Theor.}, 48(29):293001, July 2015.
\newblock \href {https://doi.org/10.1088/1751-8113/48/29/293001}
  {\path{doi:10.1088/1751-8113/48/29/293001}}.

\bibitem{cao:2025}
M.~Cao and P.~H. Diamond.
\newblock Physics of {{Edge-Core Coupling}} by {{Inward Turbulence
  Propagation}}.
\newblock {\em Phys. Rev. Lett.}, 134(23):235101, June 2025.
\newblock \href {https://doi.org/10.1103/PhysRevLett.134.235101}
  {\path{doi:10.1103/PhysRevLett.134.235101}}.

\bibitem{diamond:2025}
P.~H. Diamond, Y.~Kosuga, P.~L. Guillon, and {\"O}.~D. G{\"u}rcan.
\newblock Flux {{Jamming}}, {{Phase Transitions}} and {{Layering}} in
  {{Turbulent Magnetized Plasma}}.
\newblock {\em accepted to Philos. Trans. R. Soc. A}, October 2025.
\newblock \href {https://doi.org/10.48550/arXiv.2510.14280}
  {\path{doi:10.48550/arXiv.2510.14280}}.

\bibitem{panico:2025}
O.~Panico, Y.~Sarazin, P.~Hennequin, {\"O}.~D. G{\"u}rcan, R.~Bigu{\'e},
  G.~{Dif-Pradalier}, X.~Garbet, P.~Ghendrih, R.~Varennes, and L.~Vermare.
\newblock On the importance of flux-driven turbulence regime to address tokamak
  plasma edge dynamics.
\newblock {\em Journal of Plasma Physics}, 91(1):E26, February 2025.
\newblock \href {https://doi.org/10.1017/S0022377824001624}
  {\path{doi:10.1017/S0022377824001624}}.

\bibitem{panico:2025a}
O.~Panico, Y.~Sarazin, P.~Hennequin, {\"O}.~D. G{\"u}rcan, G.~{Dif-Pradalier},
  X.~Garbet, and R.~Varennes.
\newblock Generation of zonal flows and impact on transport in competing drift
  waves and interchange turbulence.
\newblock {\em Journal of Plasma Physics}, 91(4):E118, August 2025.
\newblock \href {https://doi.org/10.1017/S0022377825100603}
  {\path{doi:10.1017/S0022377825100603}}.

\bibitem{terry:1983}
P.~W. Terry and W.~Horton.
\newblock Drift wave turbulence in a low-order k space.
\newblock {\em The Physics of Fluids}, 26(1):106--112, January 1983.
\newblock \href {https://doi.org/10.1063/1.863997}
  {\path{doi:10.1063/1.863997}}.

\bibitem{gurcan:2022}
{\"O}.~D. G{\"u}rcan, J.~Anderson, S.~Moradi, A.~Biancalani, and P.~Morel.
\newblock Phase and amplitude evolution in the network of triadic interactions
  of the {{Hasegawa}}--{{Wakatani}} system.
\newblock {\em Physics of Plasmas}, 29(5):052306, May 2022.
\newblock \href {https://doi.org/10.1063/5.0089073}
  {\path{doi:10.1063/5.0089073}}.

\bibitem{diamond:1994}
P.~H. Diamond, Y.-M. Liang, B.~A. Carreras, and P.~W. Terry.
\newblock Self-{{Regulating Shear Flow Turbulence}}: {{A Paradigm}} for the
  {{L}} to {{H Transition}}.
\newblock {\em Phys. Rev. Lett.}, 72(16):2565--2568, April 1994.
\newblock \href {https://doi.org/10.1103/PhysRevLett.72.2565}
  {\path{doi:10.1103/PhysRevLett.72.2565}}.

\bibitem{kim:2003}
E.-j. Kim and P.~H. Diamond.
\newblock Zonal {{Flows}} and {{Transient Dynamics}} of the {{L-H Transition}}.
\newblock {\em Phys. Rev. Lett.}, 90(18):185006, May 2003.
\newblock \href {https://doi.org/10.1103/PhysRevLett.90.185006}
  {\path{doi:10.1103/PhysRevLett.90.185006}}.

\bibitem{araki:2023}
R.~Araki, W.~J.~T. Bos, and S.~Goto.
\newblock Minimal model of quasi-cyclic behaviour in turbulence driven by
  {{Taylor}}--{{Green}} forcing.
\newblock {\em Fluid Dyn. Res.}, 55(3):035507, July 2023.
\newblock \href {https://doi.org/10.1088/1873-7005/acdff7}
  {\path{doi:10.1088/1873-7005/acdff7}}.

\bibitem{kosuga:2025}
Y.~Kosuga and K.~Itoh.
\newblock Subcritical {{Bifurcation}} and {{Hysteresis}} of {{Collisionless
  Toroidal Magnetized Plasma Turbulence}}.
\newblock {\em J. Phys. Soc. Jpn.}, 94(9):094501, September 2025.
\newblock \href {https://doi.org/10.7566/JPSJ.94.094501}
  {\path{doi:10.7566/JPSJ.94.094501}}.

\bibitem{horton:1999}
W.~Horton.
\newblock Drift waves and transport.
\newblock {\em Rev. Mod. Phys.}, 71(3):735--778, April 1999.
\newblock \href {https://doi.org/10.1103/RevModPhys.71.735}
  {\path{doi:10.1103/RevModPhys.71.735}}.

\bibitem{sarazin:2021}
Y.~Sarazin, G.~{Dif-Pradalier}, X.~Garbet, P.~Ghendrih, A.~Berger, C.~Gillot,
  V.~Grandgirard, K.~Obrejan, R.~Varennes, L.~Vermare, and
  T.~{Cartier-Michaud}.
\newblock Key impact of phase dynamics and diamagnetic drive on {{Reynolds}}
  stress in magnetic fusion plasmas.
\newblock {\em Plasma Phys. Control. Fusion}, 63(6):064007, May 2021.
\newblock \href {https://doi.org/10.1088/1361-6587/abf673}
  {\path{doi:10.1088/1361-6587/abf673}}.

\bibitem{marston:2016}
J.~B. Marston, G.~P. Chini, and S.~M. Tobias.
\newblock Generalized {{Quasilinear Approximation}}: {{Application}} to {{Zonal
  Jets}}.
\newblock {\em Phys. Rev. Lett.}, 116(21):214501, May 2016.
\newblock \href {https://doi.org/10.1103/PhysRevLett.116.214501}
  {\path{doi:10.1103/PhysRevLett.116.214501}}.

\bibitem{nivarti:2024a}
G.~V. Nivarti, J.~B. Marston, and S.~M. Tobias.
\newblock Direct statistical simulation using generalised cumulant expansions.
\newblock {\em Journal of Fluid Mechanics}, 1001:A44, December 2024.
\newblock \href {https://doi.org/10.1017/jfm.2024.750}
  {\path{doi:10.1017/jfm.2024.750}}.

\bibitem{gurcan:2023}
{\"O}.~D. G{\"u}rcan.
\newblock Wave-number space networks in plasma turbulence.
\newblock {\em Rev. Mod. Plasma Phys.}, 7(1):20, May 2023.
\newblock \href {https://doi.org/10.1007/s41614-023-00122-7}
  {\path{doi:10.1007/s41614-023-00122-7}}.

\bibitem{sagaut:2006}
P.~Sagaut.
\newblock {\em Large {{Eddy Simulation}} for {{Incompressible Flows}}}.
\newblock Scientific {{Computation}}. Springer-Verlag, Berlin/Heidelberg, 2006.
\newblock \href {https://doi.org/10.1007/b137536} {\path{doi:10.1007/b137536}}.

\bibitem{lesieur:2008}
M.~Lesieur.
\newblock {\em Turbulence in {{Fluids}}}, volume~84 of {\em Fluid {{Mechanics}}
  and Its {{Applications}}}.
\newblock Springer Netherlands, Dordrecht, 2008.
\newblock \href {https://doi.org/10.1007/978-1-4020-6435-7}
  {\path{doi:10.1007/978-1-4020-6435-7}}.

\bibitem{krommes:2000}
J.~A. Krommes and C.-B. Kim.
\newblock Interactions of disparate scales in drift-wave turbulence.
\newblock {\em Phys. Rev. E}, 62(6):8508--8539, December 2000.
\newblock \href {https://doi.org/10.1103/PhysRevE.62.8508}
  {\path{doi:10.1103/PhysRevE.62.8508}}.

\bibitem{germano:1991}
M.~Germano, U.~Piomelli, P.~Moin, and W.~H. Cabot.
\newblock A dynamic subgrid-scale eddy viscosity model.
\newblock {\em Physics of Fluids A: Fluid Dynamics}, 3(7):1760--1765, July
  1991.
\newblock \href {https://doi.org/10.1063/1.857955}
  {\path{doi:10.1063/1.857955}}.

\bibitem{morel:2012}
P.~Morel, A.~Ba{\~n}{\'o}n~Navarro, M.~{Albrecht-Marc}, D.~Carati, F.~Merz,
  T.~G{\"o}rler, and F.~Jenko.
\newblock Dynamic procedure for filtered gyrokinetic simulations.
\newblock {\em Phys. Plasmas}, 19(1):012311, January 2012.
\newblock \href {https://doi.org/10.1063/1.3677366}
  {\path{doi:10.1063/1.3677366}}.

\bibitem{guillon:2025a}
P.~L. Guillon.
\newblock Penalised {{Flux-Driven Algorithm}} for {{Reduced Models}}.
\newblock \url{https://github.com/piergui/P-FLARE}, 2025.

\bibitem{guillon:2025c}
P.~L. Guillon, {\"O}.~D. G{\"u}rcan, G.~{Dif-Pradalier}, Y.~Sarazin, and
  N.~Fedorczak.
\newblock Flux-driven turbulent transport using penalisation in the
  {{Hasegawa}}--{{Wakatani}} system.
\newblock {\em Journal of Plasma Physics}, 91(5):E145, October 2025.
\newblock \href {https://doi.org/10.1017/S0022377825100895}
  {\path{doi:10.1017/S0022377825100895}}.

\bibitem{angot:1999}
P.~Angot, C.-H. Bruneau, and P.~Fabrie.
\newblock A penalization method to take into account obstacles in
  incompressible viscous flows.
\newblock {\em Numer. Math.}, 81(4):497--520, February 1999.
\newblock \href {https://doi.org/10.1007/s002110050401}
  {\path{doi:10.1007/s002110050401}}.

\bibitem{dimits:2000}
A.~M. Dimits, B.~I. Cohen, N.~Mattor, W.~M. Nevins, D.~E. Shumaker, S.~E.
  Parker, and C.~Kim.
\newblock Simulation of ion temperature gradient turbulence in tokamaks.
\newblock {\em Nucl. Fusion}, 40(3Y):661, March 2000.
\newblock \href {https://doi.org/10.1088/0029-5515/40/3Y/329}
  {\path{doi:10.1088/0029-5515/40/3Y/329}}.

\bibitem{schmitz:2012}
L.~Schmitz, L.~Zeng, T.~L. Rhodes, J.~C. Hillesheim, E.~J. Doyle, R.~J.
  Groebner, W.~A. Peebles, K.~H. Burrell, and G.~Wang.
\newblock Role of {{Zonal Flow Predator-Prey Oscillations}} in {{Triggering}}
  the {{Transition}} to {{H-Mode Confinement}}.
\newblock {\em Phys. Rev. Lett.}, 108(15):155002, April 2012.
\newblock \href {https://doi.org/10.1103/PhysRevLett.108.155002}
  {\path{doi:10.1103/PhysRevLett.108.155002}}.

\bibitem{hillesheim:2016}
J.~C. Hillesheim, E.~Delabie, H.~Meyer, C.~F. Maggi, L.~Meneses, E.~Poli, and
  {JET Contributors}.
\newblock Stationary {{Zonal Flows}} during the {{Formation}} of the {{Edge
  Transport Barrier}} in the {{JET Tokamak}}.
\newblock {\em Phys. Rev. Lett.}, 116(6):065002, February 2016.
\newblock \href {https://doi.org/10.1103/PhysRevLett.116.065002}
  {\path{doi:10.1103/PhysRevLett.116.065002}}.

\bibitem{cavedon:2016}
M.~Cavedon, T.~P{\"u}tterich, E.~Viezzer, G.~Birkenmeier, T.~Happel, F.~M.
  Laggner, P.~Manz, F.~Ryter, U.~Stroth, and T.~A.~U. Team.
\newblock Interplay between turbulence, neoclassical and zonal flows during the
  transition from low to high confinement mode at {{ASDEX Upgrade}}.
\newblock {\em Nucl. Fusion}, 57(1):014002, October 2016.
\newblock \href {https://doi.org/10.1088/0029-5515/57/1/014002}
  {\path{doi:10.1088/0029-5515/57/1/014002}}.

\bibitem{an:2017}
C.-Y. An, B.~Min, and C.-B. Kim.
\newblock Contributions of the cross phase to the plasma transport.
\newblock {\em Plasma Phys. Control. Fusion}, 59(11):115006, September 2017.
\newblock \href {https://doi.org/10.1088/1361-6587/aa8839}
  {\path{doi:10.1088/1361-6587/aa8839}}.

\bibitem{leconte:2021}
M.~Leconte and T.~Kobayashi.
\newblock Zonal profile corrugations and staircase formation: {{Role}} of the
  transport crossphase.
\newblock {\em Physics of Plasmas}, 28(1):014503, January 2021.
\newblock \href {https://doi.org/10.1063/5.0030018}
  {\path{doi:10.1063/5.0030018}}.

\bibitem{grander:2024}
F.~Grander, F.~F. Locker, and A.~Kendl.
\newblock Hysteresis in the gyrofluid resistive drift wave turbulence to zonal
  flow transition.
\newblock {\em Physics of Plasmas}, 31(5):052301, May 2024.
\newblock \href {https://doi.org/10.1063/5.0202720}
  {\path{doi:10.1063/5.0202720}}.

\bibitem{gravier:2017}
E.~Gravier, M.~Lesur, T.~Reveille, T.~Drouot, and J.~M{\'e}dina.
\newblock Transport hysteresis and zonal flow stimulation in magnetized
  plasmas.
\newblock {\em Nucl. Fusion}, 57(12):124001, October 2017.
\newblock \href {https://doi.org/10.1088/1741-4326/aa8c4c}
  {\path{doi:10.1088/1741-4326/aa8c4c}}.

\bibitem{marquant:2026}
L.~N. Marquant, P.~Morel, and {\"O}.~D. G{\"u}rcan.
\newblock On phase transition of {{ITG}} turbulence in the {{Dimits}} shift.
\newblock May 2026.
\newblock \href {https://doi.org/10.48550/arXiv.2605.26765}
  {\path{doi:10.48550/arXiv.2605.26765}}.

\bibitem{peeters:2016}
A.~G. Peeters, F.~Rath, R.~Buchholz, Y.~Camenen, J.~Candy, F.~J. Casson, S.~R.
  Grosshauser, W.~A. Hornsby, D.~Strintzi, and A.~Weikl.
\newblock Gradient-driven flux-tube simulations of ion temperature gradient
  turbulence close to the non-linear threshold.
\newblock {\em Phys. Plasmas}, 23(8):082517, August 2016.
\newblock \href {https://doi.org/10.1063/1.4961231}
  {\path{doi:10.1063/1.4961231}}.

\bibitem{stober:2000}
J.~Stober, O.~Gruber, A.~Kallenbach, V.~Mertens, F.~Ryter, A.~St{\"a}bler,
  W.~Suttrop, W.~Treutterer, and t.~A.~U. Team.
\newblock Effects of triangularity on confinement, density limit and profile
  stiffness of {{H-modes}} on {{ASDEX}} upgrade.
\newblock {\em Plasma Phys. Control. Fusion}, 42(5A):A211, May 2000.
\newblock \href {https://doi.org/10.1088/0741-3335/42/5A/324}
  {\path{doi:10.1088/0741-3335/42/5A/324}}.

\bibitem{wolf:2003}
R.~C. Wolf, Y.~Baranov, X.~Garbet, N.~Hawkes, A.~G. Peeters, C.~Challis,
  M.~de~Baar, C.~Giroud, E.~Joffrin, M.~Mantsinen, D.~Mazon, H.~Meister,
  W.~Suttrop, K.-D. Zastrow, t.~A.~U. {team}, and c.~t. t. E.-J. Workprogramme.
\newblock Characterization of ion heat conduction in {{JET}} and {{ASDEX
  Upgrade}} plasmas with and without internal transport barriers.
\newblock {\em Plasma Phys. Control. Fusion}, 45(9):1757, August 2003.
\newblock \href {https://doi.org/10.1088/0741-3335/45/9/313}
  {\path{doi:10.1088/0741-3335/45/9/313}}.

\bibitem{garbet:2004}
X.~Garbet, P.~Mantica, F.~Ryter, G.~Cordey, F.~Imbeaux, C.~Sozzi, A.~Manini,
  E.~Asp, V.~Parail, R.~Wolf, and t.~J.~E. Contributors.
\newblock Profile stiffness and global confinement.
\newblock {\em Plasma Phys. Control. Fusion}, 46(9):1351, July 2004.
\newblock \href {https://doi.org/10.1088/0741-3335/46/9/002}
  {\path{doi:10.1088/0741-3335/46/9/002}}.

\bibitem{mantica:2009}
P.~Mantica, D.~Strintzi, T.~Tala, C.~Giroud, T.~Johnson, H.~Leggate, E.~Lerche,
  T.~Loarer, A.~G. Peeters, A.~Salmi, S.~Sharapov, D.~Van~Eester, P.~C. {de
  Vries}, L.~Zabeo, and K.-D. Zastrow.
\newblock Experimental {{Study}} of the {{Ion Critical-Gradient Length}} and
  {{Stiffness Level}} and the {{Impact}} of {{Rotation}} in the {{JET
  Tokamak}}.
\newblock {\em Phys. Rev. Lett.}, 102(17):175002, April 2009.
\newblock \href {https://doi.org/10.1103/PhysRevLett.102.175002}
  {\path{doi:10.1103/PhysRevLett.102.175002}}.

\bibitem{ryter:2011}
F.~Ryter, C.~Angioni, C.~Giroud, A.~Peeters, T.~Biewer, R.~Bilato, E.~Joffrin,
  T.~Johnson, H.~Leggate, E.~Lerche, G.~Madison, P.~Mantica, D.~Van~Eester,
  I.~Voitsekhovitch, and J.~E.~T. Contributors.
\newblock Simultaneous analysis of ion and electron heat transport by power
  modulation in {{JET}}.
\newblock {\em Nucl. Fusion}, 51(11):113016, October 2011.
\newblock \href {https://doi.org/10.1088/0029-5515/51/11/113016}
  {\path{doi:10.1088/0029-5515/51/11/113016}}.

\bibitem{gurcan:2004}
{\"O}.~D. G{\"u}rcan and P.~H. Diamond.
\newblock Streamer formation and collapse in electron temperature gradient
  driven turbulence.
\newblock {\em Phys. Plasmas}, 11(2):572--583, February 2004.
\newblock \href {https://doi.org/10.1063/1.1637920}
  {\path{doi:10.1063/1.1637920}}.

\bibitem{gurcan:2004a}
{\"O}.~D. G{\"u}rcan and P.~H. Diamond.
\newblock Nonlinear elongation of two-dimensional structures in electron
  temperature gradient driven turbulence.
\newblock {\em Phys. Plasmas}, 11(11):4973--4982, November 2004.
\newblock \href {https://doi.org/10.1063/1.1786941}
  {\path{doi:10.1063/1.1786941}}.

\bibitem{camargo:1995}
S.~J. Camargo, D.~Biskamp, and B.~D. Scott.
\newblock Resistive drift-wave turbulence.
\newblock {\em Phys. Plasmas}, 2(1):48--62, January 1995.
\newblock \href {https://doi.org/10.1063/1.871116}
  {\path{doi:10.1063/1.871116}}.

\bibitem{hu:1997}
G.~Hu, J.~A. Krommes, and J.~C. Bowman.
\newblock Statistical theory of resistive drift-wave turbulence and transport.
\newblock {\em Physics of Plasmas}, 4(6):2116--2133, June 1997.
\newblock \href {https://doi.org/10.1063/1.872377}
  {\path{doi:10.1063/1.872377}}.

\bibitem{chone:2015}
L.~Ch{\^o}n{\'e}, P.~Beyer, Y.~Sarazin, G.~Fuhr, C.~Bourdelle, and S.~Benkadda.
\newblock Mechanisms and dynamics of the external transport barrier formation
  in non-linear plasma edge simulations.
\newblock {\em Nucl. Fusion}, 55(7):073010, June 2015.
\newblock \href {https://doi.org/10.1088/0029-5515/55/7/073010}
  {\path{doi:10.1088/0029-5515/55/7/073010}}.

\bibitem{gurcan:2024}
{\"O}.~D. G{\"u}rcan.
\newblock Internally {{Driven}} {$\beta$}-plane {{Plasma Turbulence Using}} the
  {{Hasegawa-Wakatani System}}, March 2024.
\newblock \href {https://doi.org/10.48550/arXiv.2403.09911}
  {\path{doi:10.48550/arXiv.2403.09911}}.

\bibitem{vallis:1993}
G.~K. Vallis and M.~E. Maltrud.
\newblock Generation of {{Mean Flows}} and {{Jets}} on a {{Beta Plane}} and
  over {{Topography}}.
\newblock {\em Journal of Physical Oceanography}, 23(7):1346--1362, July 1993.
\newblock \href
  {https://doi.org/10.1175/1520-0485(1993)023<1346:GOMFAJ>2.0.CO;2}
  {\path{doi:10.1175/1520-0485(1993)023<1346:GOMFAJ>2.0.CO;2}}.

\bibitem{basu:2003}
R.~Basu, T.~Jessen, V.~Naulin, and J.~J. Rasmussen.
\newblock Turbulent flux and the diffusion of passive tracers in electrostatic
  turbulence.
\newblock {\em Phys. Plasmas}, 10(7):2696--2703, July 2003.
\newblock \href {https://doi.org/10.1063/1.1578075}
  {\path{doi:10.1063/1.1578075}}.

\bibitem{carreras:1996a}
B.~A. Carreras, C.~Hidalgo, E.~S{\'a}nchez, M.~A. Pedrosa, R.~Balb{\'i}n,
  I.~Garc{\'i}a-Cort{\'e}s, B.~{van Milligen}, D.~E. Newman, and V.~E. Lynch.
\newblock Fluctuation-induced flux at the plasma edge in toroidal devices.
\newblock {\em Phys. Plasmas}, 3(7):2664--2672, July 1996.
\newblock \href {https://doi.org/10.1063/1.871523}
  {\path{doi:10.1063/1.871523}}.

\bibitem{garcia:2012}
O.~E. Garcia.
\newblock Stochastic {{Modeling}} of {{Intermittent Scrape-Off Layer Plasma
  Fluctuations}}.
\newblock {\em Phys. Rev. Lett.}, 108(26):265001, June 2012.
\newblock \href {https://doi.org/10.1103/PhysRevLett.108.265001}
  {\path{doi:10.1103/PhysRevLett.108.265001}}.

\bibitem{:m}
Exponentially modified {{Gaussian}} distribution.
\newblock
  \url{https://en.wikipedia.org/w/index.php?title=Exponentially_modified_Gaussian_distribution}.

\bibitem{dyson:1998a}
N.~Dyson.
\newblock {\em Chromatographic {{Integration Methods}}}.
\newblock The Royal Society of Chemistry, May 1998.
\newblock \href {https://doi.org/10.1039/9781847550514}
  {\path{doi:10.1039/9781847550514}}.

\bibitem{naulin:1999}
V.~Naulin, A.~H. Nielsen, and J.~J. Rasmussen.
\newblock Dispersion of ideal particles in a two-dimensional model of
  electrostatic turbulence.
\newblock {\em Phys. Plasmas}, 6(12):4575--4585, December 1999.
\newblock \href {https://doi.org/10.1063/1.873745}
  {\path{doi:10.1063/1.873745}}.

\bibitem{krommes:2002}
J.~A. Krommes.
\newblock Fundamental statistical descriptions of plasma turbulence in magnetic
  fields.
\newblock {\em Physics Reports}, 360(1):1--352, April 2002.
\newblock \href {https://doi.org/10.1016/S0370-1573(01)00066-7}
  {\path{doi:10.1016/S0370-1573(01)00066-7}}.

\bibitem{gurcan:2026}
{\"O}.~D. G{\"u}rcan and L.~Manfredini.
\newblock Convolutions on partially regular recurrent lattices.
\newblock {\em Communications in Nonlinear Science and Numerical Simulation},
  152:109262, January 2026.
\newblock \href {https://doi.org/10.1016/j.cnsns.2025.109262}
  {\path{doi:10.1016/j.cnsns.2025.109262}}.

\bibitem{farge:1992}
M.~Farge.
\newblock {Wavelet Transforms and their Applications to Turbulence}.
\newblock {\em Annual Review of Fluid Mechanics}, 24(Volume 24, 1992):395--458,
  January 1992.
\newblock \href {https://doi.org/10.1146/annurev.fl.24.010192.002143}
  {\path{doi:10.1146/annurev.fl.24.010192.002143}}.

\end{thebibliography}
\end{document}